\documentclass[aps,prd,longbibliography,reprint,twocolumn,amsmath,amssymb,amsfonts,showpacs,footnote]{revtex4-1}

\usepackage{ifpdf}
\usepackage{dcolumn}
\usepackage{enumitem}
\usepackage{bm}
\usepackage{here}
\usepackage{float}
\floatstyle{plaintop}
\restylefloat{table}
\ifpdf

      \usepackage[pdftex]{graphicx}  
      \usepackage[pdftex]{epsfig}

      \usepackage[pdftex]{hyperref}
\hypersetup
	{
		colorlinks,%
		citecolor=blue,%
		linkcolor=red,%
		urlcolor=blue,%
	}

\else

      \usepackage[dvips]{graphicx}  
      \usepackage[dvips]{epsfig}

      \usepackage[dvipdfm]{hyperref}
      \hypersetup
	{
		colorlinks,%
		citecolor=blue,%
		linkcolor=red,%
		urlcolor=blue,%
	}

\fi

\begin{document}

\title{Multipolar gravitational waveforms and ringdowns generated during the plunge \\ from the innermost stable circular orbit into a Schwarzschild black hole}

\author{Antoine Folacci}
\email{folacci@univ-corse.fr}
\affiliation{Equipe Physique
Th\'eorique, \\ SPE, UMR 6134 du CNRS
et de l'Universit\'e de Corse,\\
Universit\'e de Corse, Facult\'e des Sciences, BP 52, F-20250 Corte,
France}

\author{Mohamed \surname{Ould~El~Hadj}}
\email{med.ouldelhadj@gmail.com}
\affiliation{Equipe Physique
Th\'eorique, \\ SPE, UMR 6134 du CNRS
et de l'Universit\'e de Corse,\\
Universit\'e de Corse, Facult\'e des Sciences, BP 52, F-20250 Corte,
France}

\date{\today}

\begin{abstract}

We study the gravitational radiation emitted by a massive point particle plunging from slightly below the innermost stable circular orbit into a Schwarzschild black hole. We consider both even- and odd-parity perturbations and describe them using the two gauge-invariant master functions of Cunningham, Price, and Moncrief. We obtain, for arbitrary directions of observation and, in particular, outside the orbital plane of the plunging particle, the regularized multipolar waveforms, i.e., the waveforms constructed by summing over of a large number of modes, and their unregularized counterparts constructed from the quasinormal-mode spectrum. They are in excellent agreement and our results permit us to especially emphasize the impact on the distortion of the waveforms of (i) the harmonics beyond the dominant $(\ell=2,m=\pm 2)$ modes and (ii) the direction of observation, and therefore the necessity to take them into account in the analysis of the last phase of binary black hole coalescence.

\end{abstract}

\maketitle

\tableofcontents

\section{Introduction}
\label{Intr}

In this article, we shall obtain and analyze in terms of quasinormal modes (QNMs) the multipolar gravitational waveforms generated by a massive ``point particle'' plunging from slightly below the innermost stable circular orbit (ISCO) into a Schwarzschild black hole (BH). Here, it is important to note that, by multipolar waveforms, we intend waveforms constructed by superposition of a large number of modes. We shall assume an extreme mass ratio for the physical system considered, i.e., that the BH is much heavier than the particle, such a hypothesis permitting us to describe the emitted radiation in the framework of BH perturbations \cite{Regge:1957td,Zerilli:1971wd,Moncrief:1974am,Cunningham:1978zfa,Cunningham:1979px,Martel:2005ir,Nagar:2005ea}. In the context of gravitational wave physics and with the first direct gravitational-detection of a binary black hole coalescence by LIGO \cite{Abbott:2016blz}, the problem we study is of fundamental importance and there exists a large literature concerning it more or less directly
(see, e.g., Refs.~\cite{Buonanno:1998gg,
Buonanno:2000ef,Ori:2000zn,Baker:2001nu,Blanchet:2005rj,Campanelli:2006gf,Nagar:2006xv,Damour:2007xr,Sperhake:2007gu,
Mino:2008at,Hadar:2009ip,Hadar:2011vj,Price:2013paa,Hadar:2014dpa,dAmbrosi:2014llh,Hadar:2015xpa,Decanini:2015yba,Price:2015gia,DAmbrosi:2016rru}). Indeed, the ``plunge regime'' from the ISCO is the last phase of the evolution of a stellar mass object orbiting near a supermassive BH or it can be also used to describe the late-time evolution of a binary BH. Here, it is important to recall that, as a result of the radiation of gravitational waves, the eccentricity and the semimajor axis of a two-point mass system decay \cite{Peters:1964zz} and, as a consequence, for a wide class of initial conditions at large distances, the orbits are ``circularized'' and the system reaches an ISCO or an effective ISCO. Moreover, the waveform generated during the plunge regime encodes the final BH fingerprint. It should be recalled that, in this context, a multipolar description of the gravitational signal will be of fundamental interest with the enhancement of the sensitivity of laser-interferometric gravitational wave detectors (see, e.g.,  Refs.~\cite{Bernuzzi:2010ty,Bernuzzi:2010xj,Cotesta:2018fcv} and references therein). In this article, which generalize our recent work concerning the electromagnetic radiation emitted by a charged particle plunging from the ISCO into a Schwarzschild black hole \cite{Folacci:2018vtf}, we shall show, by taking into account a large number of higher harmonics, that the waveform is strongly distorted and that the distortion highly depends on the direction of observation.

It should be pointed out that our work extends the study of Hadar and Kol \cite{Hadar:2009ip} as well as the analysis of Hadar, Kol, Berti, and Cardoso in Ref.~\cite{Hadar:2011vj}. In these two articles, the multipolar and quasinormal BH responses have been theoretically constructed for any angular position of the observer but the corresponding numerical aspects are rather limited. For instance, in Ref.~\cite{Hadar:2009ip}, the authors have only plotted the quasinormal response observed in the orbital plane of the plunging particle. We go out the orbital plane in our numerical results and plots. We also go further by plotting and comparing, in the same figures and for many angular positions of the observer, the multipolar gravitational waveforms and the quasinormal ringdowns. In Refs.~\cite{Hadar:2009ip} and \cite{Hadar:2011vj}, the multipolar waveforms are never plotted and the comparison of the partial waveforms obtained in Ref.~\cite{Hadar:2011vj} with the quasinormal ringdowns constructed in Ref.~\cite{Hadar:2009ip} is achieved in an alternative way based on a numerical fitting method. Moreover, it is important to note that, due to a sign difference, our results concerning the multipolar quasinormal waveforms observed in the orbital plane of the plunging particle do not agree with the numerical results plotted in Ref.~\cite{Hadar:2009ip}.

Our paper is organized as follows. In Sec.~\ref{SecII}, after a brief overview of gravitational perturbation theory in the Schwarzschild spacetime, we establish theoretically the expression of the waveforms emitted by a massive point particle plunging from the ISCO into the BH. More precisely, we consider both the even- and odd-parity gravitational perturbations for arbitrary $(\ell,m)$ modes and describe them using the two gauge-invariant master functions of Cunningham, Price, and Moncrief \cite{Moncrief:1974am,Cunningham:1978zfa,Cunningham:1979px,Martel:2005ir,Nagar:2005ea}. We then solve, in the frequency domain and by using standard Green's function techniques, the Regge-Wheeler equation \cite{Regge:1957td} governing the odd-perturbations as well as the Zerilli-Moncrief equation \cite{Zerilli:1971wd,Moncrief:1974am} governing the even-perturbations. This is achieved after having constructed the sources for these two equations from the closed-form expression of the plunge trajectory. In Sec.~\ref{SecIII}, we extract from the results of Sec.~\ref{SecII} the QNM counterpart of the waveforms corresponding to the gravitational ringing (or ringdown) of the BH. We gather all our numerical results and their analysis in Sec.~\ref{SecIV} where we display the regularized multipolar waveforms emitted [i.e., the waveforms constructed by summing over of a large number of $(\ell,m)$ modes] and compare them with their unregularized counterparts constructed solely from the QNM spectrum. Both are obtained for arbitrary directions of observation and, in particular, outside the orbital plane of the plunging particle. It should be noted that, in the late phase of the signals, they are in excellent agreement. Moreover, our results especially emphasize the impact on the distortion of the waveforms of the harmonics beyond the dominant $(\ell=2,m=\pm 2)$ modes and of the direction of observation. At the end of this Section, we also compare our results with those displayed in Ref.~\cite{Hadar:2009ip} and propose an explanation for the disagreement found. In the Conclusion, we briefly summarize the main results obtained in this article and, in an Appendix, we carefully examine the regularization of the partial amplitudes from both the theoretical and numerical point of view. Indeed, the exact waveforms theoretically obtained in Sec.~\ref{SecII} are integrals over the radial Schwarzschild coordinate which are strongly divergent near the ISCO. For even as well as for odd perturbations, they can be numerically regularized by using the Levin's algorithm \cite{Levin1996} but only after having reduced the degree of divergence of these integrals by a succession of integrations by parts, i.e., by extending the method we developed in our work concerning the charged particle plunging from the ISCO into a Schwarzschild black hole where we encountered a similar problem \cite{Folacci:2018vtf}

Throughout this article, we adopt units such that $G = c = 1$ and we use the geometrical conventions of Ref.~\cite{Misner:1974qy}.

\section{Gravitational waves generated by the plunging massive particle}
\label{SecII}

In this section, we shall obtain theoretically the expression of the even- and odd-parity waveforms emitted by a massive point particle plunging from slightly below the ISCO into the BH. This will be achieved by working in the frequency domain and using the standard Green's function techniques. Moreover, we shall fix the notations and conventions used throughout the whole article.

\subsection{The Schwarzschild BH and the plunging massive particle}

We recall that the exterior of the Schwarzschild BH of
mass $M$ is defined by the metric
\begin{equation}\label{Metric_Schwarzschild}
ds^2= -f(r) dt^2+ f(r)^{-1}dr^2+ r^2 d\sigma_2^2
\end{equation}
where $f(r)=(1-2M/r)$ and $d\sigma_2^2=d\theta^2 + \sin^2 \theta d\varphi^2$ denotes the
metric on the unit $2$-sphere $S^2$ and with the Schwarzschild
coordinates $(t,r,\theta,\varphi)$ which satisfy $t \in ]-\infty,
+\infty[$, $r \in ]2M,+\infty[$, $\theta \in [0,\pi]$ and $\varphi
\in [0,2\pi]$. In the following, we shall also use the so-called tortoise coordinate $r_\ast \in ]-\infty,+\infty[$ defined in terms of the radial Schwarzschild coordinate $r$ by $dr/dr_\ast=f(r)$ and given by $r_\ast(r)=r+2M \ln[r/(2M)-1]$. We recall that the function $r_\ast=r_\ast(r)$
provides a bijection from $]2M,+\infty[$ to $]-\infty,+\infty[$.

\begin{figure}[h!]
\centering
\includegraphics[scale=0.6]{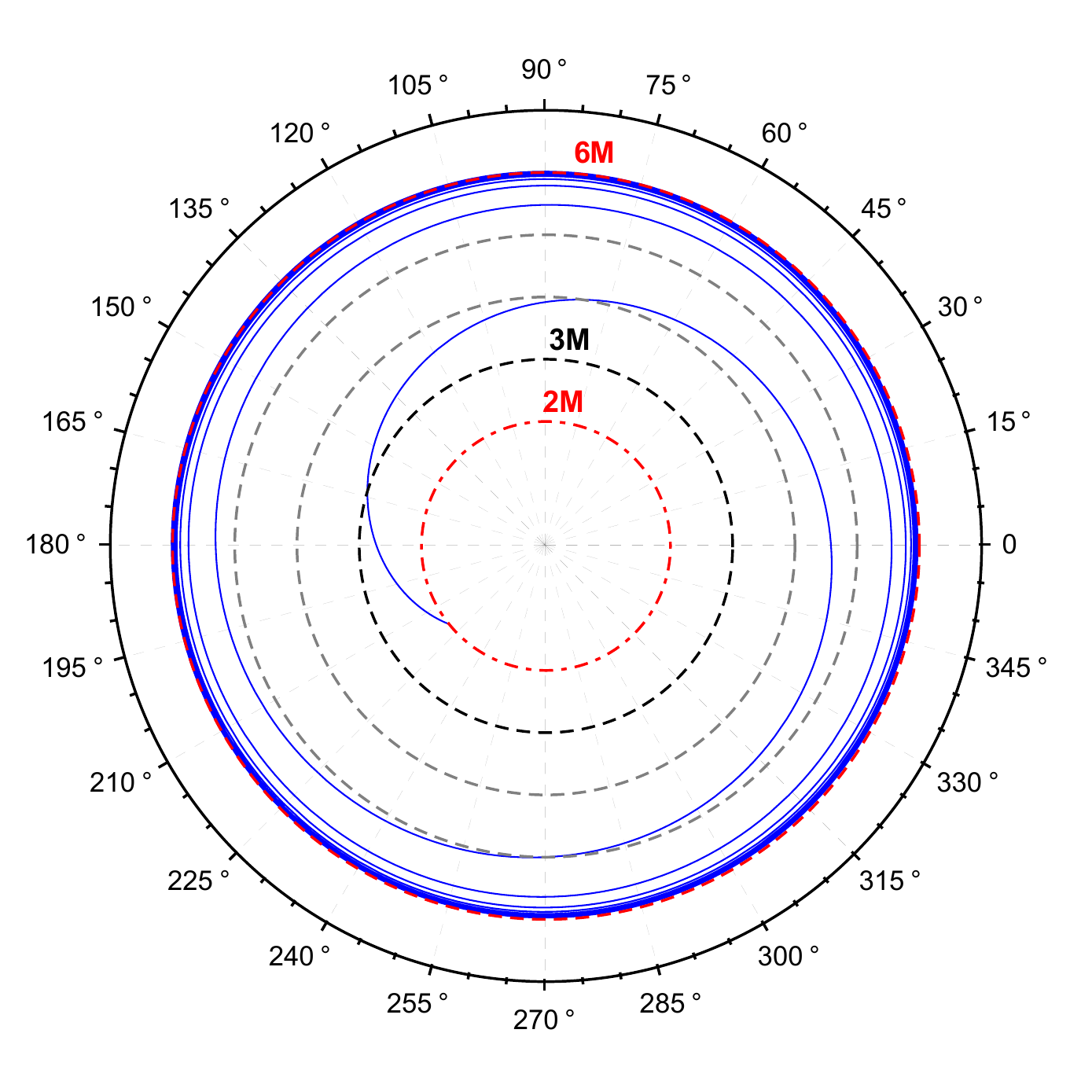}
\setlength\abovecaptionskip{-0.5ex}
\caption{\label{Trajectory_Plung} The plunge trajectory obtained from Eq.~(\ref{trajectory_plung_phi_bis}). Here, we assume that the particle starts at $r=r_\text{\tiny{ISCO}}(1-\epsilon)$ with $\epsilon=10^{-3}$ and we take $\varphi_{0}=0$. The red dashed line at $r=6M$ and the red dot-dashed line at $r=2M$ represent the ISCO and the horizon, respectively, while the black dashed line corresponds to the photon sphere at $r=3M$.}
\end{figure}

We denote by $t_{p}(\tau)$, $r_{p}(\tau)$, $\theta_{p}(\tau)$ and $\varphi_{p}(\tau)$ the coordinates of the timelike geodesic $\gamma$ followed by the plunging particle (here $\tau$ is the proper time of the particle) and by $m_0$ its mass. Without loss of generality, we can consider that its trajectory lies in the BH equatorial plane, i.e., we assume that $\theta_{p}(\tau)=\pi/2$. The geodesic equations defining $\gamma$ are given by \cite{Chandrasekhar:1985kt}
\begin{subequations}
\label{geodesic_equations}
\begin{eqnarray}
& & f(r_{p})\frac{dt_{p}}{d\tau}=\widetilde{E}, \label{geodesic_1} \\
& & r_{p}^{2}\,\,\frac{d\varphi_{p}}{d\tau}=\widetilde{L} \label{geodesic_2}
\end{eqnarray}
\noindent and
\begin{equation}
\label{geodesic_3}
\left(\frac{dr_{p}}{d\tau}\right)^{2} +\frac{\widetilde{L}^{2}}{r_{p}^{2}}f(r_{p})-\frac{2M}{r_{p}}
=\widetilde{E}^{2}-1.
\end{equation}
\end{subequations}
\noindent Here $\widetilde{E}$ and $\widetilde{L}$ are, respectively, the energy and angular momentum per unit mass of the particle which are two conserved quantities given on the ISCO, i.e., at $r=r_\text{\tiny{ISCO}}$ with
\begin{equation} \label{r_isco}
r_\text{\tiny{ISCO}}=6M,
\end{equation}
by
\begin{equation} \label{EetL_isco}
\widetilde{E}=\frac{2\sqrt{2}}{3} \quad \mathrm{and} \quad \widetilde{L}=2\sqrt{3}M.
\end{equation}
By substituting (\ref{EetL_isco}) into the geodesic equations (\ref{geodesic_1})-(\ref{geodesic_3}), we obtain after integration
\begin{eqnarray}
\label{trajectory_plung}
\frac{t_{p}(r)}{2M}\!&=&\!\frac{2\sqrt{2}\left(r-24M\right)}{2M\left(6M/r-1\right)^{1/2}}-22\sqrt{2}
\tan^{-1}\!\!\left[\left(6M/r-1\right)^{1/2}\right]\nonumber\\
&&+ 2\tanh^{-1}\!\!\left[\frac{1}{\sqrt{2}}\left(6M/r-1\right)^{1/2}\right]+  \frac{t_{0}}{2M}
\end{eqnarray}
\noindent and
\begin{equation}
\label{trajectory_plung_phi}
\varphi_{p}(r)=-\frac{2\sqrt{3}}{\left(6M/r-1\right)^{1/2}}+\varphi_{0}
\end{equation}
\noindent where $t_{0}$ and $\varphi_{0}$ are two arbitrary integration constants. From (\ref{trajectory_plung_phi}), we can write the spatial trajectory of the plunging particle in the form
\begin{equation}
\label{trajectory_plung_phi_bis}
r_{p}(\varphi)=\frac{6M}{[1+12/(\varphi-\varphi_{0})^{2}]}.
\end{equation}
We have displayed this trajectory in Fig.~\ref{Trajectory_Plung}.

\subsection{Gravitational perturbations induced by the plunging particle}

Gravitational waves emitted from the Schwarzschild BH excited by the plunging particle can be characterized by the field $h_{\mu \nu}$ which satisfies the wave equation
\begin{eqnarray}\label{Eqs perturb grav}
& &  \Box h_{\mu \nu}-h^\rho_{\phantom{\rho} \mu;\nu \rho}
 - h^\rho_{\phantom{\rho} \nu;\mu \rho}
 +h_{;\mu \nu} \nonumber \\
 & & \qquad +  g_{\mu \nu}
 (h^{\rho \sigma}_{\phantom{\rho \sigma} ;\rho \sigma}
- \Box h)=-16 \pi T_{\mu \nu}
\end{eqnarray}
where $T_{\mu \nu}$, which is the stress-energy tensor associated with the massive particle, is given by
\begin{subequations} \label{SET_moving_tot}
\begin{eqnarray}
& & T^{\mu \nu} (x) = m_0 \int_{\gamma}d\tau\, \frac{dx^\mu_p(\tau)}{d\tau} \frac{dx^\nu_p(\tau)}{d\tau} \frac{\delta^{4}(x-x_p(\tau))}{\sqrt{-g(x)}} \\
& & \phantom{T^{\mu \nu} (x)} = m_0 \frac{dx^\mu_p}{d\tau}(r) \frac{dx^\nu_p}{d\tau} (r) \left[\frac{ dr_p}{d\tau} (r)\right]^{-1}\nonumber \\
& & \qquad \qquad \quad \times \frac{\delta[t-t_p(r)] \delta[\theta- \pi/2] \delta[\varphi-\varphi_p(r)]}{r^2 \sin \theta}. \label{SET_moving}
\end{eqnarray}
\end{subequations}
In this last equation, $t_{p}(r)$ and $\varphi_{p}(r)$ are respectively given by (\ref{trajectory_plung}) and (\ref{trajectory_plung_phi}).

The resolution of the problem defined by (\ref{Eqs perturb grav}) and (\ref{SET_moving}) and, more generally, the topic of gravitational perturbations of BHs, have been the subject of lots of works since the pioneering articles by Regge and Wheeler \cite{Regge:1957td} and Zerilli \cite{Zerilli:1971wd}.  So, because gravitational perturbations of the Schwarzschild BH are very well described in the article by Martel and Poisson \cite{Martel:2005ir} as well as in the review by Nagar and Rezzolla \cite{Nagar:2005ea}, we just briefly recall some of the results we need for our particular work. The gravitational signal emitted can be described in terms of the two gauge-invariant master functions of Cunningham, Price, and Moncrief \cite{Moncrief:1974am,Cunningham:1978zfa,Cunningham:1979px} denoted by $\psi^{(e)}_{\ell m} (t,r)$ and $\psi^{(o)}_{\ell m} (t,r)$ [here, and in the following, the symbols $(e)$ and $(o)$ are respectively associated with even (polar) and odd (axial) objects according they are of even or odd parity in the antipodal transformation on the unit $2$-sphere $S^2$] which satisfy respectively the Zerilli-Moncrief and Regge-Wheeler equations
\begin{equation} \label{ZMetRW EQ}
\left[- \frac{\partial^2}{\partial t^2} + \frac{\partial^2}{\partial
r_\ast^2} - V^{(e/o)}_\ell (r)  \right] \psi^{(e/o)}_{\ell m} (t,r) = S^{(e/o)}_{\ell m} (t,r).
\end{equation}
Here the Zerilli-Moncrief potential is given by
\begin{eqnarray} \label{pot Zerilli}
& & V_\ell^{(e)}(r)=f(r)\nonumber \\
& & \qquad \times \left[\frac{\Lambda^2(\Lambda+2) r^3+6\Lambda^2 Mr^2+36\Lambda M^2r+72M^3}{(\Lambda r+6M)^2r^3} \right]\nonumber \\
& &
\end{eqnarray}
and we have for the Regge-Wheeler potential
\begin{equation} \label{pot Regge-Wheeler}
V_\ell^{(o)}(r)=f(r)\left(\frac{\Lambda+2}{r^2}-\frac{6M}{r^3} \right).
\end{equation}
In Eqs.~(\ref{pot Zerilli}) and (\ref{pot Regge-Wheeler}), we have introduced
\begin{equation} \label{Lambda_def}
\Lambda=(\ell -1) (\ell+2) = \ell (\ell+1)-2.
\end{equation}
We note that only the functions $\psi^{(e/o)}_{\ell m} (t,r)$ with $\ell = 2, 3, 4, \dots$ and $m=-\ell, -\ell+1,\dots, +\ell$ are physically relevant for our study.

We recall that the functions $S^{(e)}_{\ell m} (t,r)$ and $S^{(o)}_{\ell m} (t,r)$ are source terms which depend on the components, in the basis of tensor spherical harmonics, of the stress-tensor inducing the perturbations of the Schwarzschild spacetime. Their expressions can be found in the review by Nagar and Rezzolla:  $S^{(e)}_{\ell m} (t,r)$ is given by Eq.~(4) of the Erratum of Ref.~\cite{Nagar:2005ea} while $S^{(o)}_{\ell m} (t,r)$ is given by Eq.~(24) of Ref.~\cite{Nagar:2005ea}). After having checked these two results, we have used them to construct the sources corresponding to the stress-energy tensor (\ref{SET_moving}). By using the orthonormalization properties of the (scalar, vector and tensor) spherical harmonics \cite{Martel:2005ir,Nagar:2005ea}, we have obtained

\begin{widetext}
\begin{eqnarray} \label{Source_TR_e}
& & S^{(e)}_{\ell m} (t,r) =\frac{8 \pi m_0 [Y^{\ell m}(\pi/2,0)]^\ast}{\sqrt{2 \pi}(\Lambda+2)(\Lambda r + 6M)} f(r) \left\{ \left[\frac{3 \Lambda -2 -\frac{64 M}{\Lambda  r +6 M} +\frac{72 M^2 \left(\Lambda +2
   -m^2\right)}{r^2} +\frac{216 M^3 \left(\Lambda +2 -2 m^2\right)}{\Lambda  r^3}}{\left(6M/r-1\right)^{3/2}} \right. \right. \nonumber \\
&& \quad \left. \left. -\frac{8}{\left(6M/r-1\right)^{5/2}} -i m \frac{4 \sqrt{3} M}{r}\left(1+\frac{8}{\left(6M/r-1\right)^3}\right)  \right]\delta\left[t-t_{p}(r)\right]  -   \frac{12 \sqrt{2}  \left(r^2+12 M^2\right)}{r \left(6M/r-1\right)^3} \delta'\left[t - t_{p}(r)\right] \right\} \exp [-i m \varphi_p(r)] \nonumber \\
& &
\end{eqnarray}
and
\begin{eqnarray} \label{Source_TR_o}
& & S^{(o)}_{\ell m} (t,r) =\frac{16 \pi m_0 [X_\varphi^{\ell m}(\pi/2,0)]^\ast}{\sqrt{2 \pi} \Lambda (\Lambda+2)} f(r) \left\{ \left[ - \frac{8 \sqrt{6} M}{r^2 \left(6M/r-1\right)^{3/2}} + \frac{36 \sqrt{6} M^2}{r^3 \left(6M/r-1\right)^{5/2}}  \right. \right.\nonumber \\
&& \quad \left.\left. + i m \frac{72 \sqrt{2} M^2}{r^3 \left(6M/r-1\right)^3} \right]\delta\left[t-t_{p}(r)\right]  +   \frac{18 \sqrt{3} M \left(r^2+12 M^2\right)}{r^3 \left(6M/r-1\right)^3} \delta'\left[t - t_{p}(r)\right]  \right\} \exp [-i m \varphi_p(r)].
\end{eqnarray}
 \end{widetext}
In the last equation, we have introduced the vector spherical harmonic
\begin{equation}\label{HSV_odd}
X_\varphi^{\ell m}= - \sin \theta \frac{\partial}{\partial \theta} Y^{\ell m}.
 \end{equation}
Furthermore, we note that the coefficients $Y^{\ell m}(\pi/2,0)$ and $X_\varphi^{\ell m}(\pi/2,0)$ appearing respectively in Eqs.~(\ref{Source_TR_e}) and (\ref{Source_TR_o}) are given by
\begin{eqnarray} \label{PourSe_Y}
& & Y^{\ell m}(\pi/2,0)=\frac{2^m}{\sqrt{\pi}}\sqrt{\frac{2\ell +1}{4\pi}\frac{(\ell-m)!}{(\ell+m)!}} \nonumber \\
& & \qquad \times \frac{\Gamma[\ell/2 + m/2 +1/2]}{\Gamma[\ell/2 - m/2 +1]} \cos\left[(\ell+m)\pi/2\right].
\end{eqnarray}
and
\begin{eqnarray} \label{PourSo_X}
& & X_\varphi^{\ell m}(\pi/2,0)=\frac{2^{m+1}}{\sqrt{\pi}}\sqrt{\frac{2\ell +1}{4\pi}\frac{(\ell-m)!}{(\ell+m)!}} \nonumber \\
& & \qquad \times \frac{\Gamma[\ell/2 + m/2 +1]}{\Gamma[\ell/2 - m/2 +1/2]} \sin\left[(\ell+m)\pi/2\right].
\end{eqnarray}
Here, it is important to remark that $Y^{\ell m}(\pi/2,0)$ and hence the source (\ref{Source_TR_e}) vanish for $\ell+m$ odd while $X_\varphi^{\ell m}(\pi/2,0)$ and hence the source (\ref{Source_TR_o}) vanish for $\ell+m$ even.

It is also important to recall that the partial amplitudes $\psi^{(e/o)}_{\ell m} (t,r)$ of Cunningham, Price and Moncrief permit us to obtain the gravitational wave amplitude observed at spatial infinity (i.e., for $r \to +\infty$). In the transverse traceless gauge \cite{Misner:1974qy}, the two circularly polarized components $(h_{+},h_{\times})$ of the emitted gravitational wave are given by \cite{Nagar:2005ea}
\begin{equation} \label{hp_hc_1}
h_{+}=h^{(e)}_{+}+h^{(o)}_{+} \qquad \mathrm{and}
\qquad h_{\times}=h^{(o)}_{\times}+h^{(o)}_{\times}
\end{equation}
with\begin{subequations} \label{hp_hc_2}
\begin{eqnarray}
& & h^{(e)}_{+}= \frac{1}{r} \sum_{\ell=2}^{+\infty}
\sum_{m=-\ell}^{+\ell} \psi_{\ell m}^{(e)} \left[2 \frac{\partial^2}{\partial \theta^2} +
\ell(\ell+1) \right]Y^{\ell m}, \label{hp_hc_2a} \\
& & h^{(e)}_{\times}= \frac{1}{r} \sum_{\ell=2}^{+\infty}
\sum_{m=-\ell}^{+\ell} \psi_{\ell m}^{(e)}
\left[\frac{2}{\sin \theta}
\left(\frac{\partial^2}{\partial \theta \partial \varphi}-
\frac{\cos \theta}{\sin \theta}\frac{\partial}{\partial \varphi}
\right) \right]Y^{\ell m}, \nonumber \\
& & \label{hp_hc_2b}\\
& & h^{(o)}_{+}=
\frac{1}{r} \sum_{\ell=2}^{+\infty} \sum_{m=-\ell}^{+\ell}
\psi_{\ell m}^{(o)} \left[\frac{2}{\sin \theta}
\left(\frac{\partial^2}{\partial \theta \partial \varphi}-
\frac{\cos \theta}{\sin \theta}\frac{\partial}{\partial \varphi}
\right) \right]Y^{\ell m}, \nonumber \\
& & \label{hp_hc_2c}\\
& & h^{(o)}_{\times}= \frac{1}{r} \sum_{\ell=2}^{+\infty}
\sum_{m=-\ell}^{+\ell} \psi_{\ell m}^{(o)}
\left[- \left(2 \frac{\partial^2}{\partial \theta^2} +
\ell(\ell+1) \right) \right]Y^{\ell m}. \label{hp_hc_2d}
\end{eqnarray}
\end{subequations}
It should be noted that, due to Eqs.~(\ref{PourSe_Y}) and (\ref{PourSo_X}) [see also the remark following these equations], we have to only consider the couples $(\ell,m)$ with $\ell+m$ even in the superpositions (\ref{hp_hc_2a}) and (\ref{hp_hc_2b}) and the couples $(\ell,m)$ with $\ell+m$ odd in the superpositions (\ref{hp_hc_2c}) and (\ref{hp_hc_2d}).

\subsection{Construction of the partial amplitudes $\psi^{(e/o)}_{\ell m} (t,r)$}

In order to solve the Zerilli-Moncrief and Regge-Wheeler equations (\ref{ZMetRW EQ}), we shall work in the frequency domain by writing
\begin{equation}\label{TF_psi}
\psi^{(e/o)}_{\ell m} (t,r) = \frac{1}{\sqrt{2\pi}} \int_{-\infty}^{+\infty} d\omega \, \psi^{(e/o)}_{\omega \ell m} (r) e^{-i\omega t} \end{equation}
and
\begin{equation}\label{TF_sources}
S^{(e/o)}_{\ell m} (t,r) = \frac{1}{\sqrt{2\pi}} \int_{-\infty}^{+\infty} d\omega \, S^{(e/o)}_{\omega \ell m} (r) e^{-i\omega t}.
\end{equation}
Then, these two wave equations reduce to
\begin{equation} \label{ZMetRW EQ_Fourier}
\left[\frac{d^2}{d
r_\ast^2} + \omega^2 - V_\ell (r)  \right] \psi^{(e/o)}_{\omega \ell m} (r) = S^{(e/o)}_{\omega \ell m} (r)
\end{equation}
where the new source terms, which are obtained from (\ref{Source_TR_e}) and (\ref{Source_TR_o}), are given by
\begin{widetext}
\begin{eqnarray} \label{Source_omR_e}
& & S^{(e)}_{\omega \ell m} (r) =\frac{8 \pi m_0 [Y^{\ell m}(\pi/2,0)]^\ast}{\sqrt{2 \pi}(\Lambda+2)(\Lambda r + 6M)} f(r) \left[ i \omega   \frac{12 \sqrt{2}  \left(r^2+12 M^2\right)}{r \left(6M/r-1\right)^3}  -i m \frac{4 \sqrt{3} M}{r}\left(1+\frac{8}{\left(6M/r-1\right)^3}\right)  \phantom{+ \frac{\frac{72 M^2 \left(\Lambda +2
   -m^2\right)}{r^2} }{\left(6M/r-1\right)^{3/2}} } \right. \nonumber \\
&& \quad\left. -\frac{8}{\left(6M/r-1\right)^{5/2}} + \frac{3 \Lambda -2 -\frac{64 M}{\Lambda  r +6 M} +\frac{72 M^2 \left(\Lambda +2
   -m^2\right)}{r^2} +\frac{216 M^3 \left(\Lambda +2 -2 m^2\right)}{\Lambda  r^3}}{\left(6M/r-1\right)^{3/2}}   \right] \exp [i (\omega t_p(r)-m \varphi_p(r))]
\end{eqnarray}
and
\begin{eqnarray} \label{Source_omR_o}
& & S^{(o)}_{\omega \ell m} (r) =\frac{16 \pi m_0 [X_\varphi^{\ell m}(\pi/2,0)]^\ast}{\sqrt{2 \pi} \Lambda (\Lambda+2)} f(r) \left[ -i \omega   \frac{18 \sqrt{3} M \left(r^2+12 M^2\right)}{r^3 \left(6M/r-1\right)^3}  + i m \frac{72 \sqrt{2} M^2}{r^3 \left(6M/r-1\right)^3}  \right. \nonumber \\
&& \quad\left. + \frac{36 \sqrt{6} M^2}{r^3 \left(6M/r-1\right)^{5/2}} - \frac{8 \sqrt{6} M}{r^2 \left(6M/r-1\right)^{3/2}}  \right] \exp [i (\omega t_p(r)-m \varphi_p(r))].
\end{eqnarray}
 \end{widetext}
We have checked that our results (\ref{Source_omR_e}) and (\ref{Source_omR_o}) are in agreement with the corresponding results obtained by Hadar and Kol in Ref.~\cite{Hadar:2009ip}. It should be however noted that we do not use the same conventions for the definition of the sources, for the Fourier transform as well as the same normalization for the partial amplitudes $\psi^{(e/o)}_{\ell m} (t,r)$. Furthermore, it seems to us that our expression for the even-parity source is much simpler than theirs. It is also important to note that, due to the relation
\begin{subequations}
\begin{equation}\label{HSs_mm}
Y^{\ell -m}=(-1)^m [Y^{\ell m}]^\ast,
 \end{equation}
we have
\begin{equation}\label{HSV_odd_mm}
X_\varphi^{\ell -m} = (-1)^m [X_\varphi^{\ell m}]^\ast,
 \end{equation}
\end{subequations}
and therefore, as a direct consequence of (\ref{HSs_mm}) and (\ref{HSV_odd_mm}), we can easily observe that
\begin{equation}\label{S_omega_mm}
 S^{(e/o)}_{\omega \ell -m} = (-1)^m [ S^{(e/o)}_{-\omega \ell m}]^\ast.
\end{equation}

The Zerilli-Moncrief and Regge-Wheeler equations (\ref{ZMetRW EQ_Fourier}) can be solved by using the machinery of Green's functions (see Ref.~\cite{MorseFeshbach1953} for generalities on this topic and, e.g., Ref.~\cite{Breuer:1974uc} for its use in the context of BH physics). We consider the Green's functions $G^{(e/o)}_{\omega\ell}(r_{*},{r}_{*}')$ defined by
\begin{equation}
\label{Green_Function_1}
\left[\frac{d^{2}}{dr_{\ast}^{2}}+\omega^{2}-V_{\ell}(r)\right]G^{(e/o)}_{\omega\ell}(r_{*},{r}_{*}')=-\delta(r_{*}-{r}_{*}')
\end{equation}
which can be written as
\begin{eqnarray}
\label{Green_Function_2}
& & G^{(e/o)}_{\omega\ell}(r_{*},{r}_{*}')=-\frac{1}{W^{(e/o)}_{\ell}(\omega)} \nonumber \\
& & \qquad \times
\left\{\,
\begin{aligned}
&\!\!\phi_{\omega\ell}^{\mathrm {in} \, {(e/o)}}(r_{*})\,\phi_{\omega\ell}^{\mathrm{up} \, {(e/o)}}({r}_{*}'),\!\!&r_{*}<{r}_{\ast}',\\
&\!\!\phi_{\omega\ell}^{\mathrm{up} \, {(e/o)}}(r_{*})\,\phi_{\omega\ell}^{\mathrm {in} \, {(e/o)}}({r}_{*}'),\!\!&r_{*}>{r}_{\ast}'.
\end{aligned}
\right.
\end{eqnarray}
\noindent Here $W^{(e/o)}_{\ell}(\omega)$ denote the Wronskians of the functions $\phi_{\omega\ell}^{\mathrm {in} \, {(e/o)}}$ and $\phi_{\omega\ell}^{\mathrm{up} \, {(e/o)}}$. These functions are linearly independent solutions of the homogeneous Zerilli-Moncrief and Regge-Wheeler equations
\begin{equation}
\label{H_ZMetRW_equation}
\left[\frac{d^{2}}{dr_{\ast}^{2}}+\omega^{2}-V^{(e/o)}_{\ell}(r)\right]\phi^{(e/o)}_{\omega\ell}= 0.
\end{equation}
The functions $\phi_{\omega\ell}^{\mathrm {in} \, {(e/o)}}$ are defined by their purely ingoing behavior at the event horizon $r=2M$ (i.e., for $r_\ast \to -\infty$)
\begin{subequations}
\label{bc_in}
\begin{equation}\label{bc_1_in}
\phi^{\mathrm {in} \, {(e/o)}}_{\omega \ell} (r)\scriptstyle{\underset{r_\ast \to -\infty}{\sim}} \displaystyle{e^{-i\omega r_\ast}}
\end{equation}
while, at spatial infinity $r \to +\infty$ (i.e., for $r_\ast \to +\infty$), they have an
asymptotic behavior of the form
\begin{equation}\label{bc_2_in}
\phi^{\mathrm {in} \, {(e/o)}}_{\omega  \ell}(r) \scriptstyle{\underset{r_\ast \to +\infty}{\sim}}
\displaystyle{A^{(-,{e/o})}_\ell (\omega) e^{-i\omega r_\ast} + A^{(+,{e/o})}_\ell (\omega) e^{+i\omega r_\ast}}.
\end{equation}
\end{subequations}
Similarly, the functions $\phi^{\mathrm{up} \, {(e/o)}}_{\omega \ell }$ are defined by their purely outgoing behavior at spatial infinity
\begin{subequations}
\label{bc_up}
\begin{equation}\label{bc_1_up}
\phi^{\mathrm{up} \, {(e/o)}} (r)\scriptstyle{\underset{r_\ast \to +\infty}{\sim}}
 \displaystyle{e^{+i\omega r_\ast}}
\end{equation}
and, at the horizon, they have an asymptotic behavior of the form
\begin{equation}\label{bc_2_up}
\phi^{\mathrm{up} \, {(e/o)}}_{\omega \ell }(r) \scriptstyle{\underset{r_\ast \to -\infty}{\sim}}\displaystyle{
B^{(-,{e/o})}_\ell (\omega) e^{-i\omega r_\ast}  + B^{(+,{e/o})}_\ell (\omega) e^{+i\omega r_\ast}}.
\end{equation}
\end{subequations}
In the previous expressions, the coefficients $A^{(-,{e/o})}_\ell (\omega)$, $A^{(+,{e/o})}_\ell (\omega)$, $B^{(-,{e/o})}_\ell (\omega)$ and $B^{(+,{e/o})}_\ell (\omega)$ are complex amplitudes. By evaluating the Wronskians $W^{(e/o)}_\ell (\omega)$ at $r_\ast \to -\infty$ and $r_\ast \to +\infty$, we obtain
\begin{equation}
\label{Well}
W^{(e/o)}_\ell (\omega) =2i\omega A^{(-,{e/o})}_\ell (\omega) = 2i\omega B^{(+,{e/o})}_\ell (\omega).
\end{equation}

Here, it is worth noting some important properties of the coefficients $A^{(\pm,{e/o})}_\ell (\omega)$ and of the functions $\phi^{\mathrm {in} \, {(e/o)}}_{\omega  \ell}(r)$ that we will use extensively later. They are a direct consequence of Eqs.~(\ref{H_ZMetRW_equation}) and (\ref{bc_in}) and they are valid whether $\omega$ is real or complex. We have
\begin{subequations}\label{PhiIN_mm}
\begin{equation}\label{PhiIN_mm_a}
\phi^{\mathrm {in} \, {(e/o)}}_{-\omega \ell}(r) = [\phi^{\mathrm {in} \, {(e/o)}}_{\omega \ell}(r)]^\ast
\end{equation}
and
\begin{equation}\label{PhiIN_mm_b}
A^{(\pm,{e/o})}_\ell (-\omega)=[A^{(\pm,{e/o})}_\ell (\omega)]^\ast.
\end{equation}
\end{subequations}
It is important to also recall that the solutions of the homogeneous Zerilli-Moncrief and Regge-Wheeler equations (\ref{H_ZMetRW_equation}) are related by the Chandrasekhar-Detweiler transformation \cite{Chandrasekhar:1975zza,Chandrasekhar:1985kt}
\begin{widetext}
\begin{equation}\label{ChandraDet_transf_PHI}
\left[\Lambda (\Lambda+2) - i (12 M\omega) \right] \phi^{(e)}_{\omega\ell} =\left[\Lambda (\Lambda+2) + \frac{72 M^2}{r (\Lambda r+6M)} f(r) + 12M f(r) \frac{d}{dr} \right] \phi^{(o)}_{\omega\ell}
\end{equation}
\end{widetext}
and, as a consequence, the coefficients $A^{(\pm,{e/o})}_\ell (\omega)$ satisfy the relations
\begin{subequations}\label{ChandraDet_transf}
\begin{equation}\label{ChandraDet_transf_a}
A^{(-,e)}_\ell (\omega)=A^{(-,o)}_\ell (\omega)
\end{equation}
and
\begin{equation}\label{ChandraDet_transf_b}
A^{(+,e)}_\ell (\omega)=\frac{\Lambda (\Lambda +2) + i  (12M\omega) }{\Lambda (\Lambda +2) - i  (12M\omega) } A^{(+,o)}_\ell (\omega).
\end{equation}
\end{subequations}

Using the Green's functions (\ref{Green_Function_2}), we can show that the solutions of the Moncrief-Zerilli and Regge-Wheeler equations with source (\ref{ZMetRW EQ_Fourier}) are given by
\begin{subequations}\label{G_Sol_ZMetRW_Eq_en_ast}
\begin{eqnarray}
& & \psi^{(e/o)}_{\omega\ell m}(r)=- \int_{-\infty}^{+\infty}d{r}_{*}'\,G^{(e/o)}_{\omega\ell}(r_{*},{r}_{*}')
S^{(e/o)}_{\omega\ell m}({r}_{*}') \label{G_Sol_ZMetRW_Eq_en_ast_a}\\
& & \phantom{\psi^{(e/o)}_{\omega\ell m}(r)} = - \int_{2M}^{6M} \frac{dr'}{f(r')}   G^{(e/o)}_{\omega\ell}(r,r')
S^{(e/o)}_{\omega\ell m}(r'). \label{G_Sol_ZMetRW_Eq_en_ast_b}
\end{eqnarray}
\end{subequations}
For $r \to +\infty$, the solutions (\ref{G_Sol_ZMetRW_Eq_en_ast_b}) reduce to the asymptotic expressions
\begin{eqnarray}
\label{Partial_Response_1}
& & \psi^{(e/o)}_{\omega\ell m}(r)= \frac{e^{+i \omega r_\ast }}{2i\omega A^{(-,{e/o})}_\ell (\omega)}  \nonumber \\
& & \qquad\quad \times \int_{2M}^{6M} \frac{dr'}{f(r')} \,\phi_{\omega\ell}^{\mathrm {in} \, {(e/o)}}(r')
\,S^{(e/o)}_{\omega\ell m}(r').
\end{eqnarray}
This result is a consequence of Eqs.~(\ref{Green_Function_2}), (\ref{bc_1_up}) and (\ref{Well}).

We can now obtain the solutions of the Zerilli-Moncrief and Regge-Wheeler equations (\ref{ZMetRW EQ}) by inserting (\ref{Partial_Response_1}) into (\ref{TF_psi}) and we have, in the time domain, for the $(\ell,m)$ waveforms
\begin{eqnarray}
\label{partial_response_def}
& & \psi^{(e/o)}_{\ell m}(t,r) = \frac{1}{\sqrt{2\pi}} \int_{-\infty}^{+\infty} d\omega  \left(\frac{e^{- i \omega [t-r_\ast (r)]}}{2 i \omega A_{\ell}^{(-,{e/o})}(\omega)}\right)\nonumber\\
& & \qquad\qquad \times\,  \int_{2M}^{6M} \frac{dr'}{f(r')} \,\phi_{\omega\ell}^{\mathrm {in} \, {(e/o)}}(r')
\,S^{(e/o)}_{\omega\ell m}(r').
\end{eqnarray}
Here it is important to note that these partial waveforms satisfy
\begin{equation}\label{Psi_t_mm}
\psi^{(e/o)}_{\ell -m} = (-1)^m [\psi^{(e/o)}_{\ell m}]^\ast.
\end{equation}
This is a direct consequence of the definition (\ref{TF_psi}) and of the relation
\begin{equation}\label{Psi_om_mm}
\psi^{(e/o)}_{\omega \ell -m} = (-1)^m [\psi^{(e/o)}_{-\omega \ell m}]^\ast
\end{equation}
which is easily obtained from (\ref{Partial_Response_1}) and (\ref{S_omega_mm}) by noting that the solutions $\phi_{\omega\ell}^{\mathrm {in} \, {(e/o)}}$ of the problem (\ref{H_ZMetRW_equation})-(\ref{bc_in}) and the associated coefficients $A_{\ell}^{(-,{e/o})}(\omega)$ satisfy the relations (\ref{PhiIN_mm}). The relations (\ref{Psi_t_mm}) and (\ref{HSs_mm}) permit us to check that the gravitational wave amplitudes (\ref{hp_hc_2}) are purely real.

\begingroup
\begin{table*}[htp]
\caption{\label{tab:table1} The first quasinormal frequencies $\omega_{\ell n}$ and the associated excitation factors $\mathcal{B}_{\ell n}^{(e/o)}$.}
\smallskip
\centering
\begin{tabular}{cccc}
\hline
\hline
 $(\ell, n)$ & $2M \omega_{\ell n}$ & $\mathcal{B}_{\ell n}^{(e)}$ & $\mathcal{B}_{\ell n}^{(o)}$
 \\ \hline
 $(2,1)$  & $0.747343 - 0.177925 i $  & $\phantom{-} 0.120928 + 0.070666 i$  & $\phantom{-} 0.126902 + 0.0203151 i$ \\
 $(3,1)$  & $1.198890 - 0.185406 i $  &  $-0.088969 - 0.061177 i$ & $-0.093890 - 0.049193 i$  \\
 $(4,1)$  & $1.618360 - 0.188328 i $  & $\phantom{-} 0.062125 + 0.069099 i$  & $\phantom{-} 0.065348 + 0.065239 i $ \\
 $(5,1)$  & $2.024590 - 0.189741 i $  & $-0.036403 - 0.074807 i$  & $-0.038446 - 0.073524 i $  \\
 $(6,1)$  & $2.424020 - 0.190532 i $  & $\phantom{-}0.011847 + 0.075196 i$   & $\phantom{-} 0.013129 + 0.074877 i $  \\
 $(7,1)$  & $2.819470 - 0.191019 i $  & $\phantom{-} 0.010478 - 0.069864 i$  & $\phantom{-} 0.009689 - 0.069924 i $ \\
 $(8,1)$  & $3.212390 - 0.191341 i $  & $-0.029331 + 0.059378 i$  & $-0.028863 + 0.059573 i $ \\
 $(9,1)$  & $3.603590 - 0.191565 i $  & $\phantom{-} 0.043628 - 0.044836 i$  & $\phantom{-} 0.043370 - 0.045060 i $  \\
 $(10,1)$ & $3.993576 - 0.191728 i $  & $-0.052616 + 0.027669 i$  & $-0.052494 + 0.027875 i $  \\
\hline
\hline
\end{tabular}
\end{table*}
\endgroup

\begin{table*}[htp]
\caption{\label{tab:table2} The excitation coefficients $\mathcal{C}_{\ell m n}^{(e/o)}$ and  $\mathcal{D}_{\ell m n}^{(e/o)}$ corresponding to the quasinormal frequencies $\omega_{\ell n}$ and the excitation factors $\mathcal{B}_{\ell n}^{(e/o)}$ of Table \ref{tab:table1}.}
\smallskip
\centering
\resizebox{\textwidth}{!}{%
\begin{tabular}{ccccc}
\hline
\hline
 $(\ell, n)$ &$\mathcal{C}_{\ell \ell n}^{(e)}$ & $\mathcal{D}_{\ell \ell n}^{(e)}$ & $\mathcal{C}_{\ell {\ell-1} n}^{(o)}$ & $\mathcal{D}_{\ell {\ell-1} n}^{(o)}$
 \\ \hline
 $(2,1)$ & $-2.2873\times10^{-5} - 1.1413\times10^{-5} i$  & $\phantom{-} 1.2331\times10^{-7} + 1.6133\times10^{-8} i$  &$\phantom{-}  1.3027 \times10^{-5} - 3.7333\times10^{-6} i $ & $\phantom{-}  1.7606\times10^{-8} - 9.7647\times10^{-7} i$   \\
 $(3,1)$ & $-3.4129\times10^{-6} + 3.7527\times10^{-7} i$  & $-8.9300\times10^{-10}-1.6665\times10^{-9} i$   & $\phantom{-} 1.5710\times10^{-6} + 2.3923\times10^{-7} i$ & $-1.1817\times10^{-8} + 4.4988\times10^{-9} i$ \\
 $(4,1)$   & $\phantom{-} 3.4146\times10^{-7} - 8.3671\times10^{-7} i $  & $\phantom{-} 5.0989\times10^{-11} + 2.4768\times10^{-11} i $  & $-2.3418\times10^{-7} + 3.0764\times10^{-7} i$ & $\phantom{-} 2.4210\times10^{-10} - 2.4821\times10^{-10} i$\\
 $(5,1)$   & $-2.5796\times10^{-9} + 3.2195\times10^{-7} i $  & $-1.4574\times10^{-12} + 1.4770\times10^{-12} i $  & $\phantom{-} 3.3738\times10^{-8} - 1.2541\times10^{-7} i$ & $\phantom{-} 8.0274\times10^-{12} + 2.2079\times10^{-11} i$  \\
 $(6,1)$   &$-2.1803\times10^{-9} - 1.3600\times10^{-7} i $   & $-1.1924\times10^{-13} + 5.2103\times10^{-14} i $   & $-1.1875\times10^{-8} + 5.0871\times10^{-8} i$ &  $-3.3844\times10^{-12} + 5.2475\times10^{-12} i$  \\
 $(7,1)$ & $-1.0692\times10^{-8} + 6.3204\times10^{-8} i $   & $\phantom{-} 4.0607\times10^{-14} - 6.8884\times10^{-14} i $   & $\phantom{-} 9.2800\times10^{-9} - 2.1679\times10^{-8} i$  & $-3.0697\times10^{-12} + 1.3254\times10^{-12} i$  \\
 $(8,1)$  & $\phantom{-} 1.5158\times10^{-8} - 2.8867\times10^{-8} i $   & $-2.9165\times10^{-14} + 3.4046\times10^{-14} i $   & $-7.5475\times10^{-9} + 8.7255\times10^{-9} i$ & $-2.0486\times10^{-12} - 3.6520\times10^{-13} i$  \\
 $(9,1)$ & $-1.3623\times10^{-8} + 1.1053\times10^{-8} i $   & $\phantom{-} 2.0391\times10^{-14} - 2.3094\times10^{-14} i $ & $\phantom{-} 5.3721\times10^{-9} - 2.6496\times10^{-9} i$  & $-1.0116\times10^{-12} - 9.2597\times10^{-13} i$ \\
 $(10,1)$& $\phantom{-} 9.6352\times10^{-9} - 2.0768\times10^{-9} i $  & $-1.4482\times10^{-14} + 1.2662\times10^{-14} i $  & $-3.2557\times10^{-9} - 1.9158\times10^{-11} i$ & $-2.6317\times10^{-13} - 9.0578\times10^{-13} i$   \\
\hline
\hline
\end{tabular}%
}
\end{table*}

\section{Quasinormal ringings due to the plunging massive particle}
\label{SecIII}

In this section, we shall construct the quasinormal ringings associated with the gravitational wave amplitudes (\ref{hp_hc_2}). Of course, they can be obtained by summing over the ringings associated with all the partial amplitudes $\psi^{(e/o)}_{\ell m}(t,r)$. In order to extract from these partial amplitudes the corresponding quasinormal ringings $\psi^{\text{\tiny{QNM}} \, (e/o)}_{\ell m n}(t,r)$, the contour of integration over $\omega$ in Eq.~(\ref{partial_response_def}) may be deformed (see, e.g., Ref.~\cite{Leaver:1986gd}). This deformation permits us to capture the zeros of the Wronskians (\ref{Well}) lying in the lower part of the complex $\omega$ plane and which are the complex frequencies $\omega_{\ell n}$ of the $(\ell,n)$ QNMs. We note that, for a given $\ell$, $n=1$ corresponds to the fundamental QNM (i.e., the least damped one) while $n=2, 3, \dots$ to the overtones. We also recall that the spectrum of the quasinormal frequencies is symmetric with respect to the imaginary axis, i.e., that if $\omega_{\ell n}$ is a quasinormal frequency lying in the fourth quadrant, $-\omega_{\ell n}^{*}$ is the symmetric quasinormal frequency lying in the third one. We then easily obtain

\begin{equation}
\label{partial_response_QNM_1}
\psi^{\text{\tiny{QNM}} \, (e/o)}_{\ell m}(t,r) = \sum^{+\infty}_{n=1} \psi^{\text{\tiny{QNM}} \, (e/o)}_{\ell m n}(t,r)
\end{equation}
with
\begin{eqnarray}
\label{partial_response_QNM_2}
 \psi^{\text{\tiny{QNM}} \, (e/o)}_{\ell m n}(t,r) = && -\sqrt{2\pi}\left({\cal{C}}^{(e/o)}_{\ell m n}\,\, e^{-i \omega_{\ell n}[t-r_\ast(r)]}\,\,\vphantom{e^{i \omega_{\ell n}^{\ast}t}}\right.\nonumber\\
&&\left.+\,\,{\cal{D}}^{(e/o)}_{\ell m n}e^{+i \omega^\ast_{\ell n}[t-r_\ast(r)]} \right).
\end{eqnarray}
\noindent In the previous expression, ${\cal{C}}^{(e/o)}_{\ell m n}$ and ${\cal{D}}^{(e/o)}_{\ell m n}$ denote the extrinsic excitation coefficients (see, e.g., Refs.~\cite{Leaver:1986gd,Berti:2006wq,Zhang:2013ksa}) which are here defined by
\begin{subequations}\label{excitation_coeffs}
\begin{equation}
\label{excitation_coeff_C}
{\cal{C}}^{(e/o)}_{\ell m n}={\cal{B}}^{(e/o)}_{\ell n} \left[\int_{2M}^{6M} \frac{dr'}{f(r')} \, \frac{\phi_{\omega \ell}^{\mathrm {in} \, {(e/o)}}(r') }{A_{\ell}^{(+,e/o)}(\omega)}S^{(e/o)}_{\omega \ell m}(r') \right]_{\omega=\omega_{\ell n}}
\end{equation}
and
\begin{eqnarray}\label{excitation_coeff_D}
& & {\cal{D}}^{(e/o)}_{\ell m n}= \left[{\cal{B}}^{(e/o)}_{\ell n}\right]^\ast \nonumber \\
& & \qquad \times \left[\int_{2M}^{6M}  \frac{dr'}{f(r')} \, \frac{\phi_{\omega \ell}^{\mathrm {in} \, {(e/o)}}(r') }{A_{\ell}^{(+,e/o)}(\omega)}S^{(e/o)}_{\omega \ell m}(r') \right]_{\omega=-\omega_{\ell n}^{*}}
\end{eqnarray}
\end{subequations}
while
\begin{equation}
\label{excitation_factor}
{\cal{B}}^{(e/o)}_{\ell n}=\left[\frac{1}{2 \omega}\,\,\frac{A_{\ell}^{(+,e/o)}(\omega)}{\frac{d}{d \omega}A_{\ell}^{(-,e/o)}(\omega)}\right]_{\omega=\omega_{\ell n}}
\end{equation}
are the even and odd excitation factors associated with the $(\ell,n)$ QNM of complex frequency $\omega_{\ell n}$. The first term in the right hand side (r.h.s.)~of Eq.~(\ref{partial_response_QNM_2}) is the contribution of the quasinormal frequency $\omega_{\ell n}$ lying in the fourth quadrant of the $\omega$ plane while the second one is the contribution of $-\omega_{\ell n}^{\ast}$, i.e., its symmetric with respect to the imaginary axis. In front of the bracket in the r.h.s.~of Eq.~(\ref{excitation_coeff_D}), the coefficients $\left[{\cal{B}}^{(e/o)}_{\ell n}\right]^\ast$ are nothing else than the even and odd excitation factors associated with the $(\ell,n)$ QNM of complex frequency $-\omega_{\ell n}^{\ast}$. They are obtained from (\ref{excitation_factor}) by using the properties (\ref{PhiIN_mm_b}). A few remarks are in order:

\begin{enumerate}[label=(\arabic*)]

\item Thanks to Chandrasekhar and Detweiler, we know that the zeros of the Wronskians (\ref{Well}), i.e., the quasinormal frequencies $\omega_{\ell n}$ (and $-\omega_{\ell n}^{\ast}$), do not depend on the parity sector. This is a consequence of the relation (\ref{ChandraDet_transf_a}).

   \item The excitation factors (\ref{excitation_factor}) depend on the parity sector because their expressions involve the coefficients $A^{(+,{e/o})}_\ell (\omega)$ which are parity dependent [see Eq.~(\ref{ChandraDet_transf_b})]. It is interesting to recall that this was not the case for the problem of the electromagnetic field generated by charged particle plunging into the Schwarzschild BH \cite{Folacci:2018vtf}. By combining (\ref{ChandraDet_transf}) with the definition (\ref{excitation_factor}), we obtain

       \begin{equation}\label{ExcitationF_ChandraDet_transf}
{\cal{B}}^{(e)}_{\ell n}=\frac{\Lambda (\Lambda +2) + i  (12M \omega_{\ell n}) }{\Lambda (\Lambda +2) - i  (12M \omega_{\ell n}) } {\cal{B}}^{(o)}_{\ell n}.
\end{equation}

       \item The excitation coefficients  (\ref{excitation_coeff_C}) and  (\ref{excitation_coeff_D}) depend on the parity sector because they are constructed from the excitation factors as well as from wave equations with sources which are parity dependent [see Eq.~(\ref{ZMetRW EQ_Fourier})].

   \item  In our problem, the spherical symmetry of the Schwarzschild BH is broken due to the asymmetric plunging trajectory. It is this dissymmetry which, in connection with the presence of the azimuthal number $m$, forbids us to gather the two terms in Eq.~(\ref{partial_response_QNM_2}).

   \item  It is however important to note that the excitation coefficients (\ref{excitation_coeffs}) are related by
\begin{equation}
\label{excitation_coeffs_CetD_prop}
{\cal{C}}^{(e/o)}_{\ell -m n}= (-1)^m \left[ {\cal{D}}^{(e/o)}_{\ell m n} \right]^\ast
\end{equation}
[this is due to the properties (\ref{PhiIN_mm})] and hence that the quasinormal waveforms (\ref{partial_response_QNM_1}) satisfy
\begin{equation}\label{PsiQNM_t_mm}
\psi^{\text{\tiny{QNM}} \, (e/o)}_{\ell -m} = (-1)^m [\psi^{\text{\tiny{QNM}} \, (e/o)}_{\ell m}]^\ast.
\end{equation}
The quasinormal gravitational amplitudes obtained from (\ref{hp_hc_2}) by replacing $\psi^{(e/o)}_{\ell m}(t,r)$ with $\psi^{\text{\tiny{QNM}} \, (e/o)}_{\ell m}(t,r)$ are then purely real as a consequence of the relations (\ref{PsiQNM_t_mm}) and (\ref{HSs_mm}).

       \item The ringing amplitudes $\psi^{\text{\tiny{QNM}} \, (e/o)}_{\ell m n}(t,r)$ and $\psi^{\text{\tiny{QNM}} \, (e/o)}_{\ell m}(t,r)$ do not provide physically relevant results at ``early times'' due to their  exponentially divergent behavior as $t$ decreases. It is necessary to determine, from physical considerations (see below), the time beyond which these quasinormal waveforms can be used, i.e., the starting time $t_\mathrm{start}$ of the BH ringing.
\end{enumerate}

\section{Multipolar waveforms and quasinormal ringdowns}
\label{SecIV}

\begin{figure*}
\centering
 \includegraphics[scale=0.55]{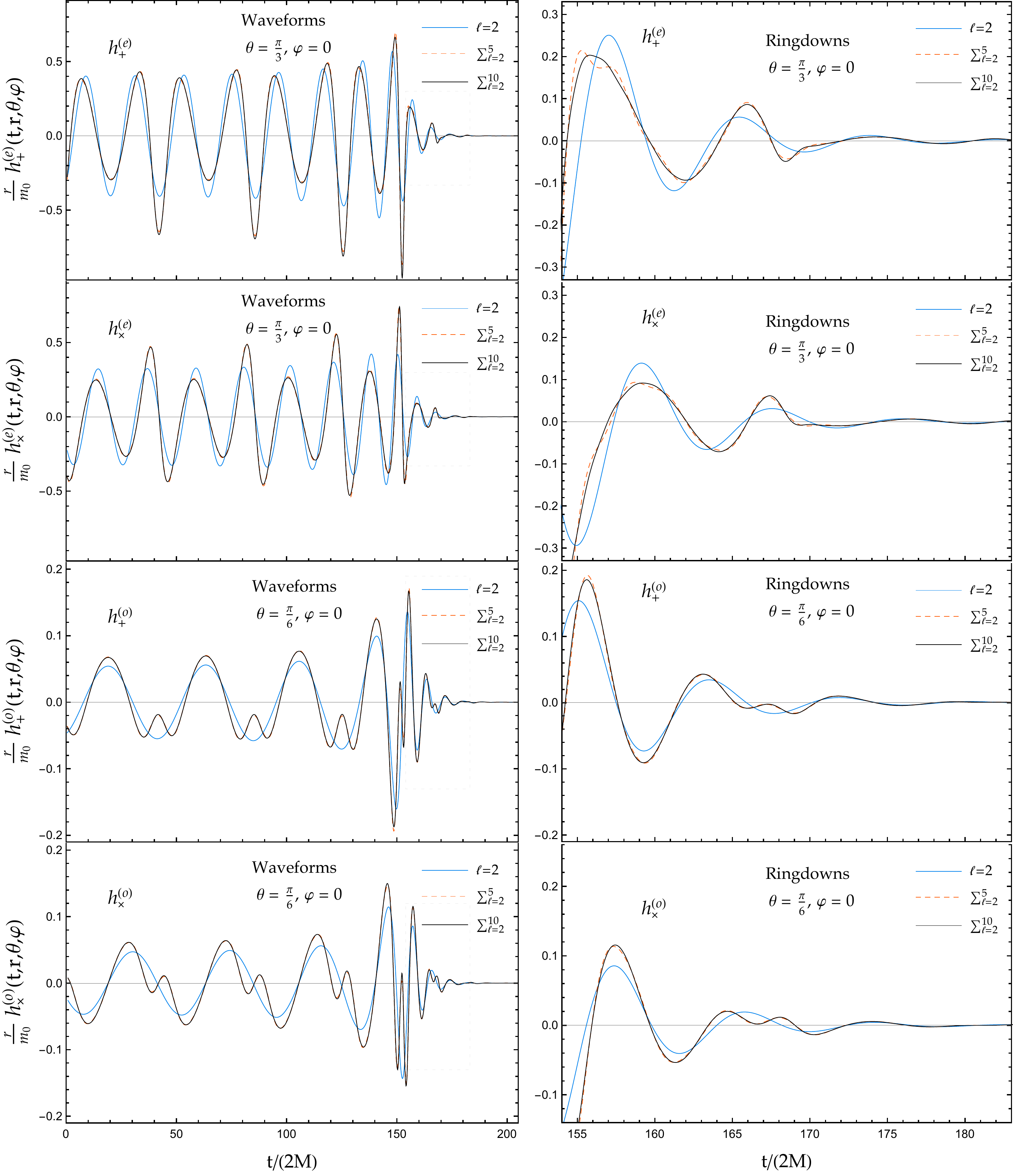}
\caption{\label{h_sum_Ringdowns_sum_ell} Components $h_{+/\times}^{(e/o)}$ of the gravitational wave observed at infinity in the direction $(\theta= \pi/3 ,\varphi=0)$ (for even components) and $(\theta= \pi/6 ,\varphi=0)$ (for odd components). We emphasize the impact of the harmonics beyond the dominant $(\ell=2,m= \pm 2)$ modes on the multipolar waveforms and their ringdowns (zoom in on the waveforms).}
\end{figure*}

\begin{figure*}[ht]
\centering
 \includegraphics[scale=0.53]{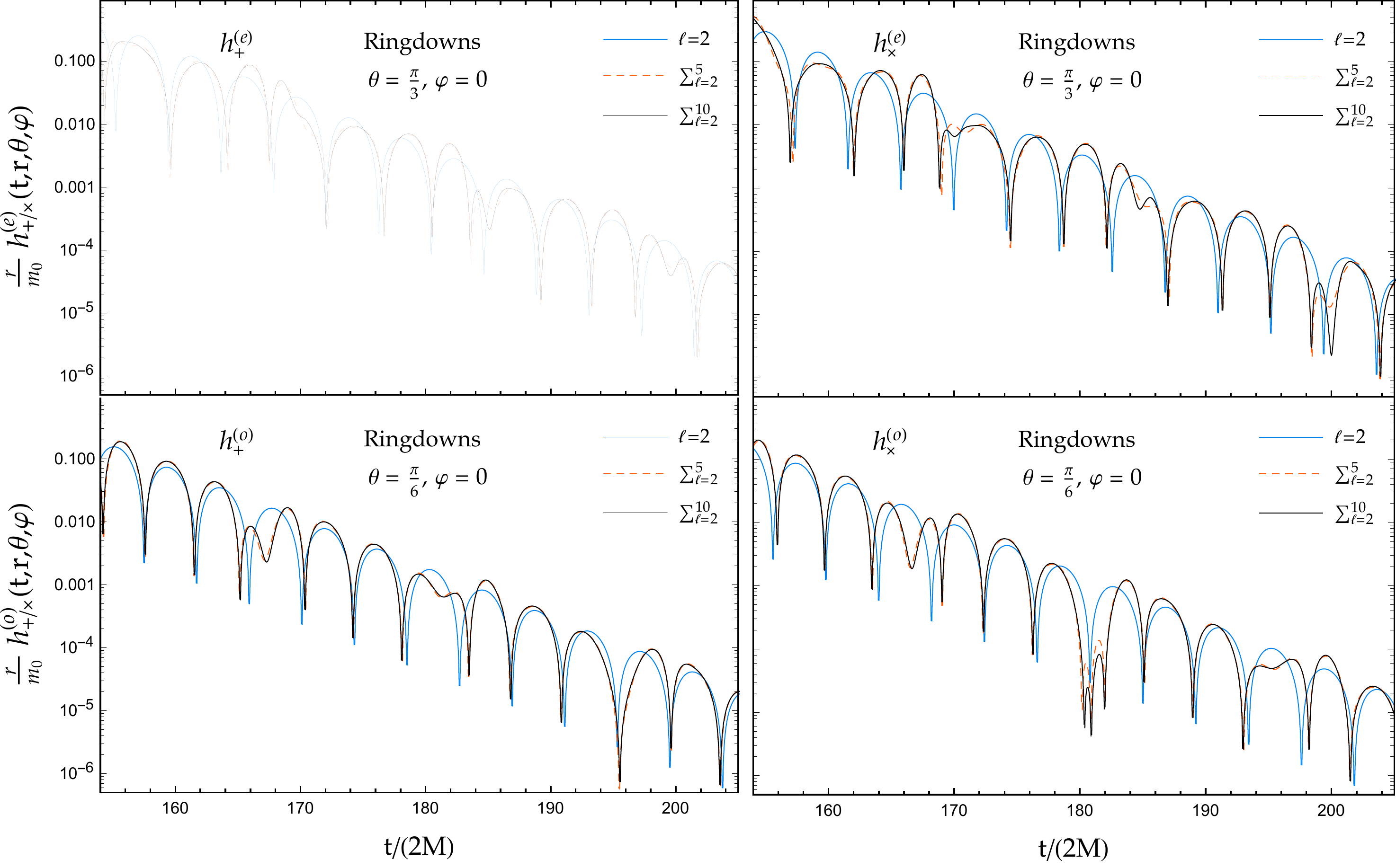}
\caption{\label{h_sum_Ringdowns_sum_ell_log} Complement to Fig.~\ref{h_sum_Ringdowns_sum_ell}. Semi-log graphs highlighting the impact on the ringdowns of the harmonics beyond the dominant $(\ell=2,m= \pm 2)$ modes.}
\end{figure*}

\begin{figure*}
\centering
 \includegraphics[scale=0.5]{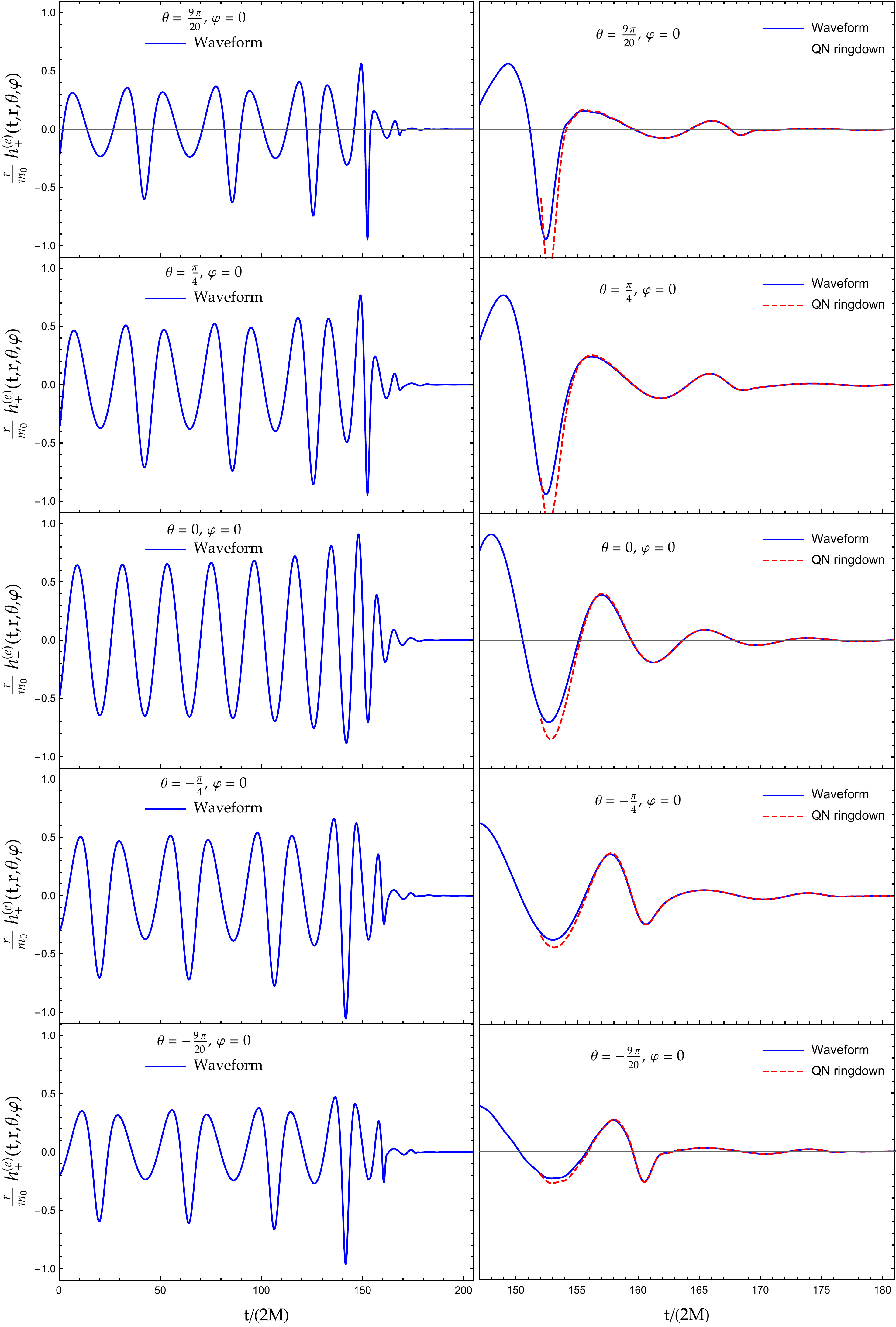}
\caption{\label{hp_even} Multipolar gravitational waveforms $h_{+}^{(e)}$ observed at infinity for various directions above the orbital plane of the plunging particle. We consider $\varphi=0$ and we study the distortion of the multipolar waveform and of the associated quasinormal ringdown when $\theta$ varies between $-\pi/2$  and $+\pi/2$. We note that, for $\theta=0$, only the $(\ell=2,m=\pm 2)$ modes contribute to the signal.}
\end{figure*}

\begin{figure*}
\centering
 \includegraphics[scale=0.5]{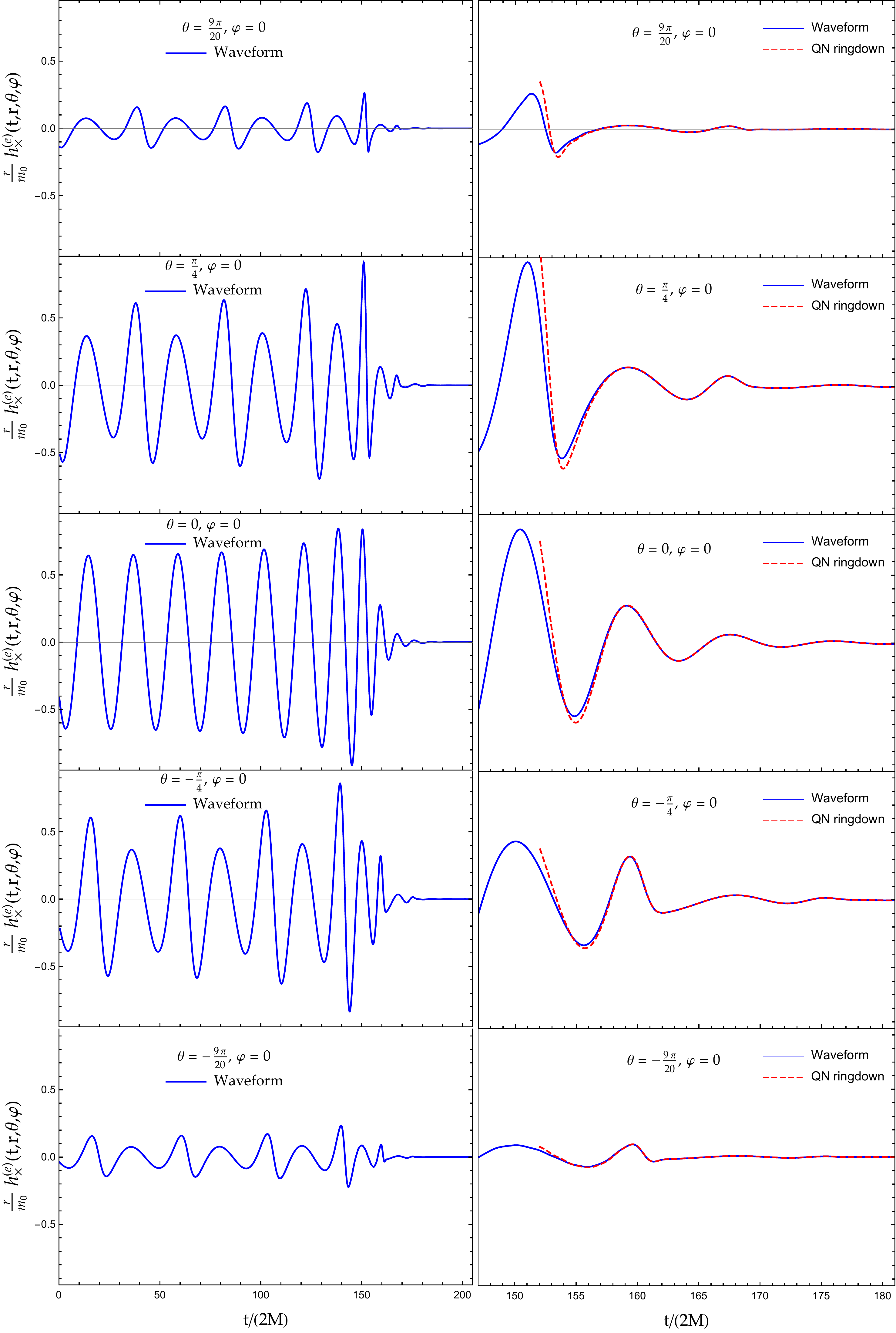}
\caption{\label{hx_even} Multipolar gravitational waveforms $h_{\times}^{(e)}$ observed at infinity for various directions above the orbital plane of the plunging particle. We consider $\varphi=0$ and we study the distortion of the multipolar waveform and of the associated quasinormal ringdown when $\theta$ varies between $-\pi/2$  and $+\pi/2$. We note that $h_{\times}^{(e)}$ vanishes for $\theta = \pm \pi/2$ and that, for $\theta=0$, only the $(\ell=2,m=\pm 2)$ modes contribute to the signal.}
\end{figure*}

\begin{figure*}
\centering
 \includegraphics[scale=0.50]{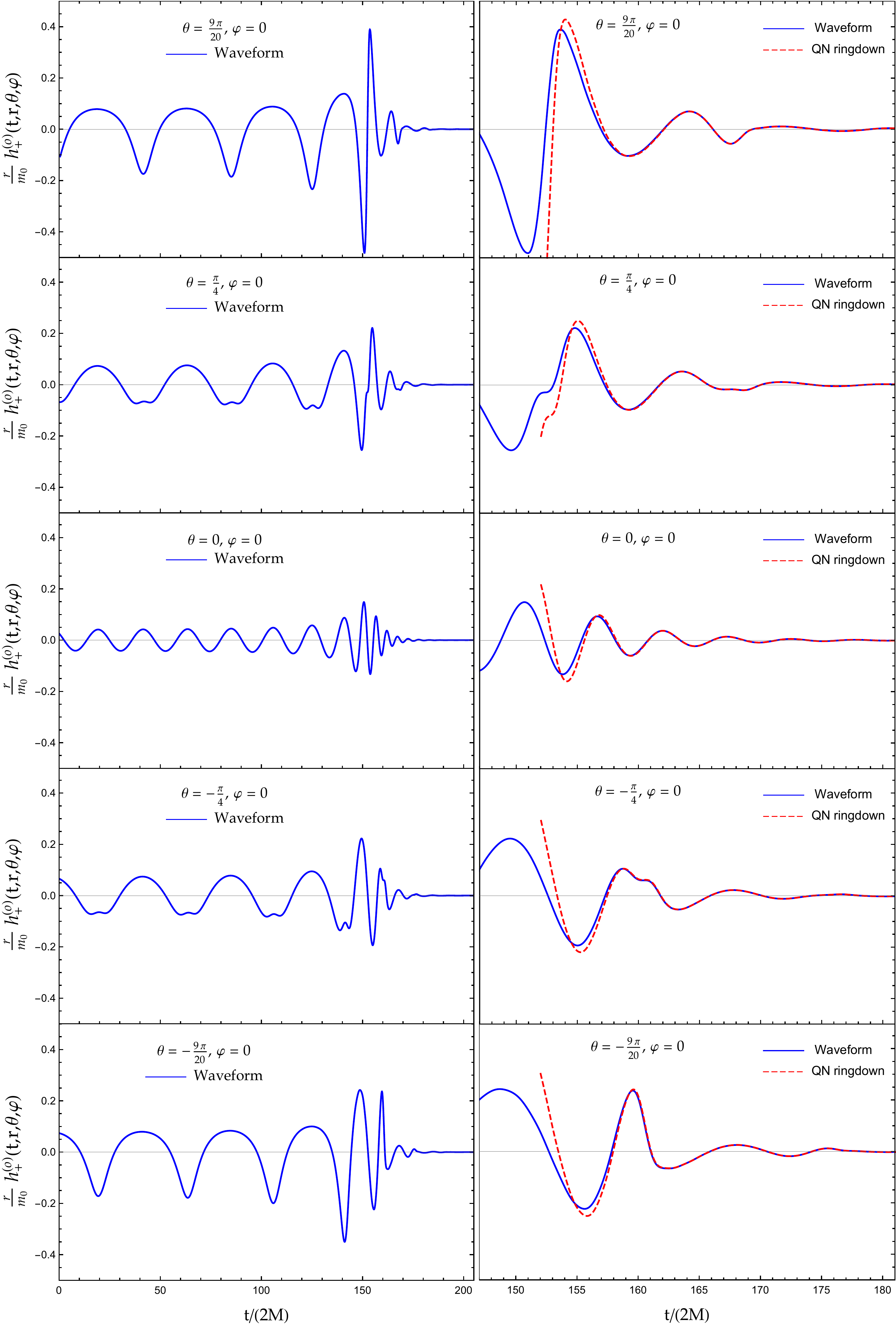}
\caption{\label{hp_odd} Multipolar gravitational waveforms  $h_{+}^{(o)}$ observed at infinity for various directions above the orbital plane of the plunging particle. We consider $\varphi=0$ and we study the distortion of the multipolar waveform and of the associated quasinormal ringdown when $\theta$ varies between $-\pi/2$  and $+\pi/2$. We note that, for $\theta=0$, only the $(\ell=3,m=\pm 2)$ modes contribute to the signal.}
\end{figure*}

\begin{figure*}
\centering
 \includegraphics[scale=0.50]{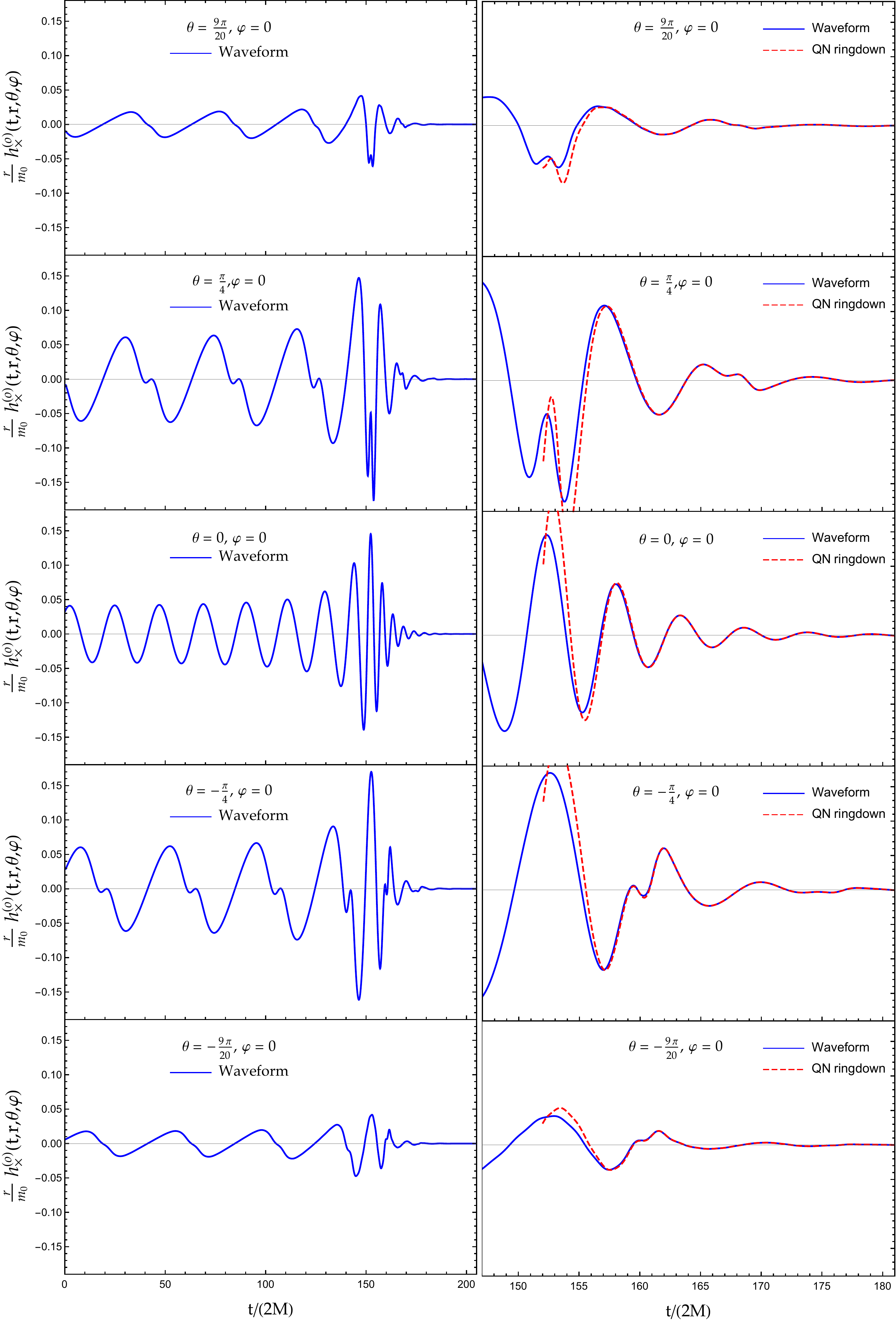}
\caption{\label{hx_odd} Multipolar gravitational waveforms  $h_{\times}^{(o)}$ observed at infinity for various directions above the orbital plane of the plunging particle. We consider $\varphi=0$ and we study the distortion of the multipolar waveform and of the associated quasinormal ringdown when $\theta$ varies between $-\pi/2$  and $+\pi/2$. We note that $h_{\times}^{(o)}$ vanishes for $\theta = \pm \pi/2$ and that, for $\theta=0$, only the $(\ell=3,m=\pm 2)$ modes contribute to the signal.}
\end{figure*}

\begin{figure*}[ht]
\centering
 \includegraphics[scale=0.53]{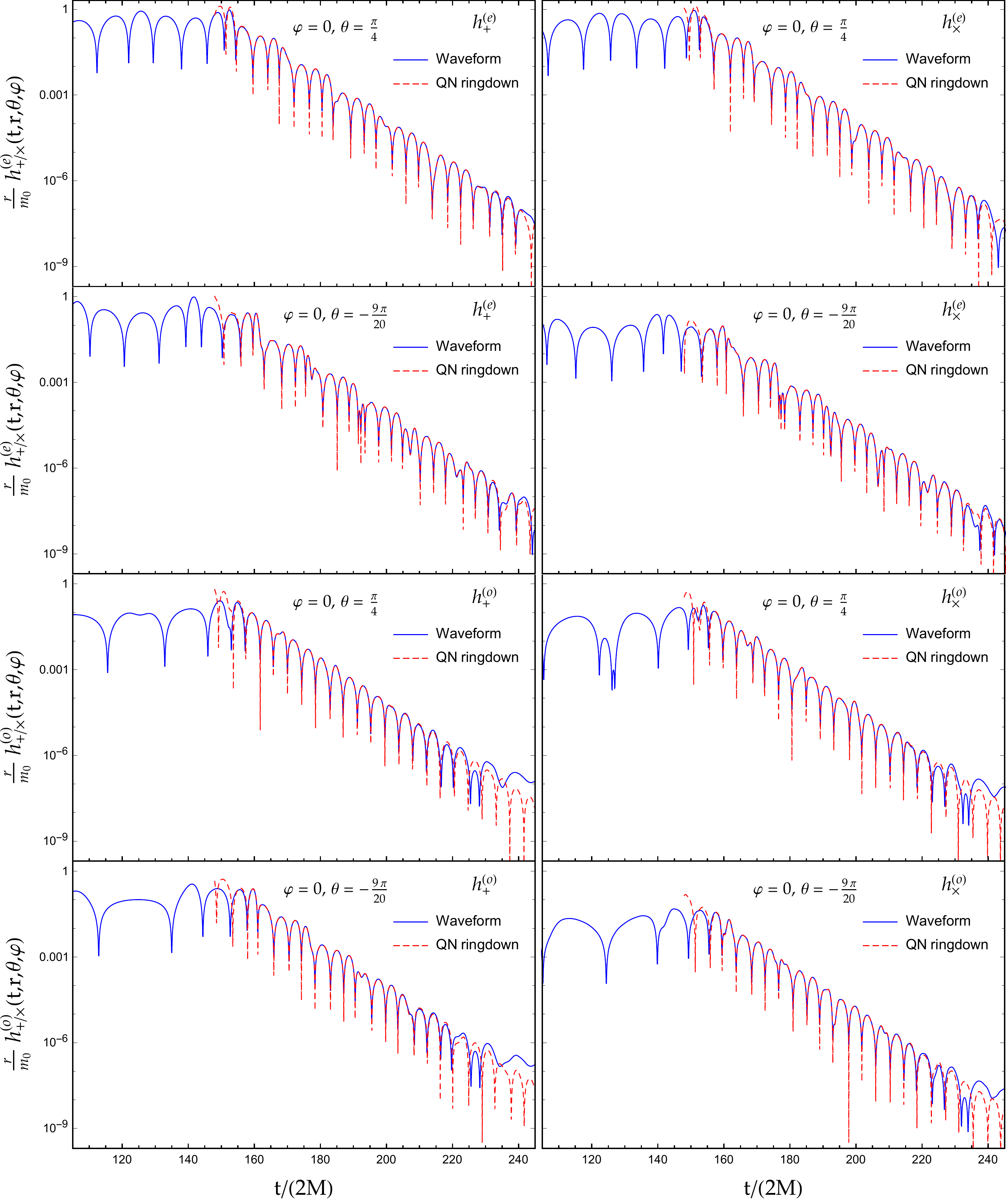}
\caption{\label{h_log} Semi-log graphs of some multipolar waveforms showing the dominance of the quasinormal ringing at intermediate times and the agreement of the regularized waveforms with the unregularized quasinormal responses.}
\end{figure*}

\subsection{Numerical methods}
\label{SecIVa}

In order to construct the gravitational wave amplitudes (\ref{hp_hc_2}), it is first necessary to obtain numerically the partial amplitudes $\psi^{(e/o)}_{\ell m}(t,r)$ given by (\ref{partial_response_def}). For that purpose, using {\it Mathematica} \cite{Mathematica}:
\begin{enumerate}[label=(\arabic*)]
   \item We have determined the functions  $\phi^{\mathrm{in} \, (e/o)}_{\omega \ell}$ as well as the coefficients $A^{(-,e/o)}_\ell (\omega)$. This has been achieved by integrating numerically the homogeneous Zerilli-Moncrief and Regge-Wheeler equations (\ref{H_ZMetRW_equation}) with the Runge-Kutta method.  We have initialized the process with Taylor series expansions converging near the horizon and we have compared the solutions to asymptotic expansions with ingoing and outgoing behavior at spatial infinity that we have decoded by Pad\'e summation. Our numerical calculations have been performed independently for the two parities and we have checked their robustness and internal consistency by using the relations (\ref{ChandraDet_transf_PHI}) and (\ref{ChandraDet_transf}).

   \item  We have regularized the partial amplitudes $\psi^{(e/o)}_{\omega \ell m}(r)$ given by (\ref{Partial_Response_1}), i.e., the Fourier transform of the partial amplitudes $\psi^{(e/o)}_{\ell m}(t,r)$. Indeed, these amplitudes as integrals over the radial Schwarzschild coordinate are strongly divergent near the ISCO. This is due to the behavior of the sources (\ref{Source_omR_e}) and (\ref{Source_omR_o}) in the limit $r \to 6M$. The regularization process is described in the Appendix. It consists in replacing the partial amplitudes (\ref{Partial_Response_1}) by their counterparts (\ref{REG_RES}) and to evaluate the result by using Levin's algorithm \cite{Levin1996}.

      \item We have Fourier transformed $\psi^{(e/o)}_{\omega \ell m}(r)$ to get the final result.

\end{enumerate}
\noindent Then, from the partial amplitudes $\psi^{(e/o)}_{\ell m}(t,r)$, it is possible to obtain the components $h_{+}^{(e/o)}$ and $h_{\times}^{(e/o)}$ of the gravitational signal by using the sums (\ref{hp_hc_2}). We have constructed the even components from the $(\ell,m)$ modes with $\ell=2,\dots,10$ and $m=\pm \ell$ which constitute the main contributions. Similarly, we have constructed the odd components from the $(\ell,m)$ modes with $\ell=2,\dots,10$ and $m=\pm (\ell -1)$. In fact, it is not necessary to take higher values for $\ell$ because, in general, they do not really modify the numerical sums (\ref{hp_hc_2}) (see also the discussion in Sec.~\ref{R_and_S}).

In order to construct the quasinormal ringings associated with the gravitational wave amplitudes (\ref{hp_hc_2}), it is necessary to obtain numerically the partial amplitudes $\psi^{\text{\tiny{QNM}} \, (e/o)}_{\ell m}(t,r)$ given by (\ref{partial_response_QNM_1}) and, as a consequence, we need the quasinormal frequencies $\omega_{\ell n}$, the excitation factors ${\cal{B}}^{(e/o)}_{\ell n}$ as well as the excitation coefficients ${\cal{C}}^{(e/o)}_{\ell m n}$ and ${\cal{D}}^{(e/o)}_{\ell m n}$. For that purpose:
\begin{enumerate}[label=(\arabic*)]
   \item We have first determined the quasinormal frequencies $\omega_{\ell n}$ by using the method developed by Leaver \cite{Leaver:1985ax}. We have implemented numerically this method by using the Hill determinant approach of Majumdar and Panchapakesan \cite{mp}.

    \item  We have then obtained the excitation factors ${\cal B}^{(e/o)}_{\ell n}$, as well as the functions $\phi_{\omega_{\ell n} \ell}^{\mathrm {in} \, {(e/o)}}(r)$ and the coefficients ${A_{\ell}^{(+,e/o)}(\omega_{\ell n})}$ by integrating numerically the homogeneous Zerilli-Moncrief and Regge-Wheeler equations (\ref{H_ZMetRW_equation}) (for $\omega=\omega_{\ell n}$) with the Runge-Kutta method and then by comparing the solutions to asymptotic expansions with ingoing and outgoing behavior at spatial infinity. Our numerical results are in agreement with the theoretical relations (\ref{ChandraDet_transf}) and (\ref{ExcitationF_ChandraDet_transf}).

     \item We have finally determined the excitation coefficients ${\cal{C}}^{(e/o)}_{\ell m n}$ and ${\cal{D}}^{(e/o)}_{\ell m n}$ by evaluating the integrals in Eqs.~(\ref{excitation_coeff_C}) and (\ref{excitation_coeff_D}). It should be noted that we do not have to regularize these integrals if we work with weakly damped QNMs. Indeed, let us consider, for example, the integrals in Eq.~(\ref{excitation_coeff_C}) which defines the excitation coefficients ${\cal{C}}^{(e/o)}_{\ell m n}$. For $\omega=\omega_{\ell n}$, due to the term $\exp[i\omega t_p(r')]$, the sources $S^{(e/o)}_{\omega \ell m}(r')$ given by (\ref{Source_omR_e}) and (\ref{Source_omR_o}) vanish exponentially in the limit $r' \to 6M$ and the integrals are convergent at the upper limit. In the limit $r' \to 2M$, as a consequence of (\ref{bc_1_in}), we have $\phi^{\mathrm{in} \, (e/o)}_{\omega \ell} (r') \propto (r'-2M)^{-i (2M\omega)}$ and, as a consequence of (\ref{trajectory_plung}), we have $t_p(r') \simeq -2M \ln [r'-2M] + \mathrm{Cte}$. Thus, the integrands are proportional to $(r'-2M)^{-i (4M\omega)}$ (see also Sec.~IVB of Ref.~\cite{Hadar:2009ip}). By noting that
\begin{eqnarray}
\label{Comportement_hor_CoefExcit}
& & \int_{2M+\epsilon} dr' \, (r'-2M)^{-i (4M\omega_{\ell n})} \nonumber \\
& & \qquad = \frac{\epsilon^{-i (4M \operatorname{Re}[\omega_{\ell n}])}}{1-i (4M\omega_{\ell n})} \epsilon^{1 + (4M \operatorname{Im}[\omega_{\ell n}])}
\end{eqnarray}
we can see that the integrals in Eq.~(\ref{excitation_coeff_C}) are convergent at the lower limit $2M$ if
\begin{equation}
\label{Condition_CV_hor_CoefExcit}
2M \operatorname{Im}[\omega_{\ell n}] > -1.
\end{equation}
Such a condition is satisfied by the QNMs we shall consider below.
\end{enumerate}
In fact, for a given $\ell$, it is possible to consider only the fundamental QNM ($n=1$) which is the least damped one. Moreover, we need only the excitation coefficients ${\cal{C}}^{(e)}_{\ell m n}$ and ${\cal{D}}^{(e)}_{\ell m n}$ with $\ell=2,\dots,10$ and $m=\pm \ell$ and the excitation coefficients ${\cal{C}}^{(o)}_{\ell m n}$ and ${\cal{D}}^{(o)}_{\ell m n}$ with $\ell=2,\dots,10$ and $m=\pm (\ell -1)$. In Tables \ref{tab:table1} and \ref{tab:table2}, we provide the various ingredients permitting us to construct the quasinormal ringings associated with the gravitational wave amplitudes (\ref{hp_hc_2}). It should be finally recalled that it is necessary to select a starting time $t_\mathrm{start}$ for the ringings. By taking $t_\mathrm{start}=t_p(3M)$, i.e., the moment the particle crosses the photon sphere, we have obtained physically relevant results.

\subsection{Results and comments}
\label{R_and_S}

In Figs.~\ref{h_sum_Ringdowns_sum_ell}-\ref{h_log}, we have considered the components $h_{+/\times}^{(e/o)}$ of the gravitational waves observed at infinity. The multipolar waveforms have been obtained by assuming that the particle starts at $r=r_\text{\tiny{ISCO}}(1-\epsilon)$ with $\epsilon=10^{-4}$ and, furthermore, in Eqs.~(\ref{trajectory_plung}) and (\ref{trajectory_plung_phi}), we have taken  $\varphi_{0}=0$ and chosen $t_{0}/(2M)$ in order to shift the interesting part of the signal in the window $t/(2M)\in[0,245]$. Without loss of generality, we have constructed only the signals for directions above the orbital plane of the plunging particle. Indeed, we could obtain those observed below that plane by using the symmetry properties of the vector spherical harmonics in the antipodal transformation on the unit $2$-sphere $S^2$. Moreover, we have assumed that the observer lies in the plane $\varphi=0$. In fact, for any other value of $\varphi$, the behavior of the signals is very similar. The results corresponding to arbitrary values of $\theta$ and $\varphi$ are available to the interested reader upon request.

In Figs.~\ref{h_sum_Ringdowns_sum_ell} and \ref{h_sum_Ringdowns_sum_ell_log}, we have focused our attention on the construction of the multipolar waveforms by summing the expressions (\ref{hp_hc_2}) over the harmonics beyond the dominant $(\ell=2,m= \pm 2)$ modes. Of course, the necessity to take higher harmonics into account clearly appears but we can also note that the sums truncated at $\ell =5$ already provide strong results.

The distortion of the multipolar waveforms and of the associated quasinormal ringdowns clearly appears in Figs.~\ref{hp_even}-\ref{hx_odd}. It can be observed in the adiabatic phase corresponding to the quasicircular motion of the particle near the ISCO (see Fig.~\ref{Trajectory_Plung}) as well as in the ringdown phase. It is due to the ``large'' number of $(\ell,m)$ modes taken into account in the sums (\ref{hp_hc_2}) and is strongly dependent on the direction of the observer.

The multipolar waveforms and the associated quasinormal ringdowns are in excellent agreement as can be seen in Figs.~\ref{hp_even}-\ref{hx_odd} or, more clearly, in Fig.~\ref{h_log} where we work with semi-log graphs. Here, it is important to recall (see Sec.~\ref{SecIVa}) that it has been necessary to regularize the former while the latter are unregularized.

\begin{figure*}
\centering
 \includegraphics[scale=0.50]{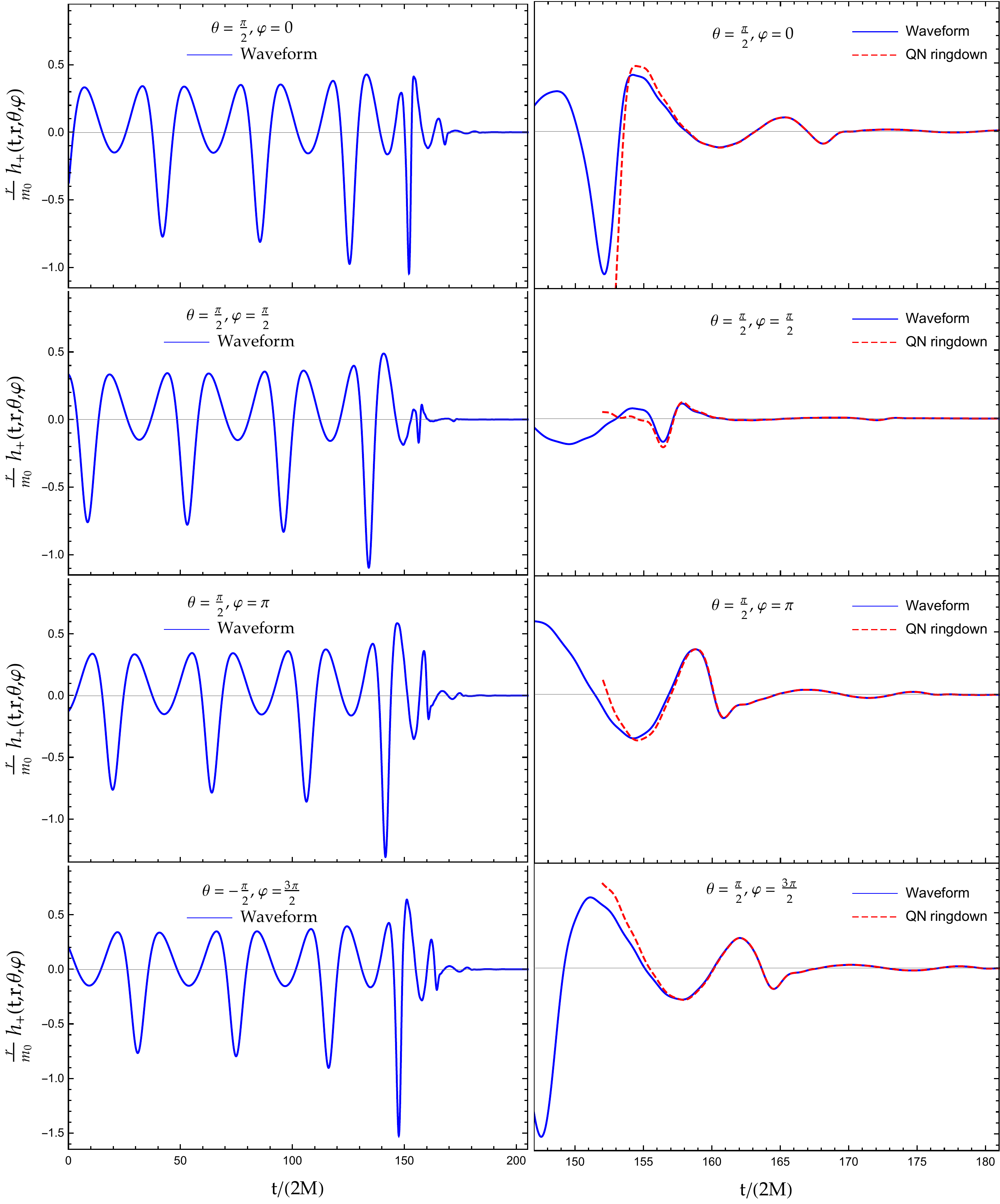}
\caption{\label{hplus_plan_equatorial} Multipolar gravitational waveforms $h_{+}=h_{+}^{(e)}+h_{+}^{(o)}$ and associated quasinormal ringdowns observed at infinity for various directions in the orbital plane of the plunging particle. The observation directions are $\varphi=0$, $\varphi=\pi/2$, $\varphi=\pi$ and $\varphi=3\pi/2$. The results displayed in the right panel do not agree with those presented in Fig.~4 of Ref.~\cite{Hadar:2009ip}.}
\end{figure*}

\begin{figure*}
\centering
 \includegraphics[scale=0.50]{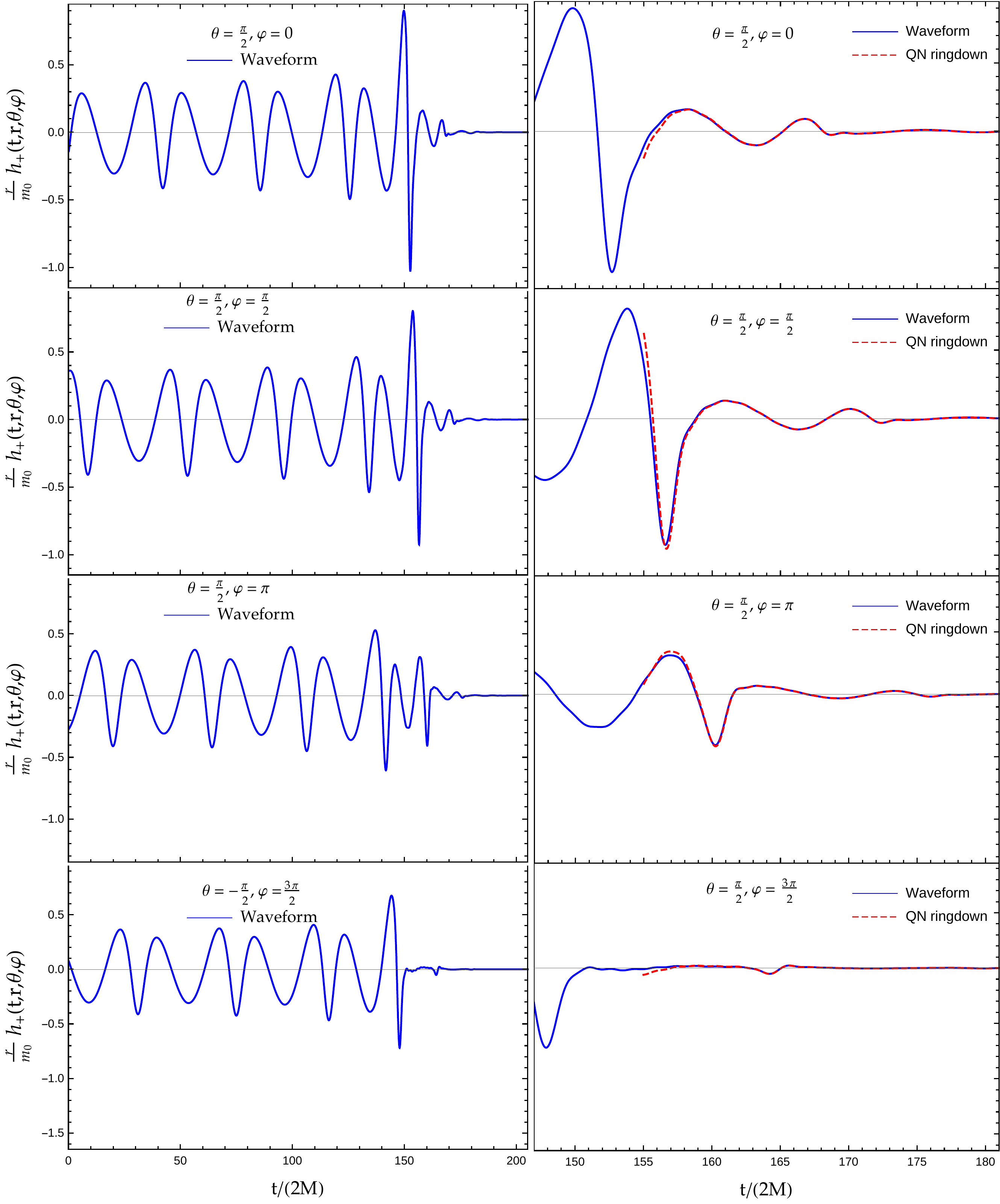}
\caption{\label{HK_hplus_plan_equatorial} Multipolar gravitational waveforms $h_{+}=h_{+}^{(e)}-h_{+}^{(o)}$ and associated quasinormal ringdowns observed at infinity for various directions in the orbital plane of the plunging particle. The observation directions are $\varphi=0$, $\varphi=\pi/2$, $\varphi=\pi$ and $\varphi=3\pi/2$. The results displayed in the right panel agree with those presented in Fig.~4 of Ref.~\cite{Hadar:2009ip}.}
\end{figure*}

Finally, in Fig.~\ref{hplus_plan_equatorial}, in order to compare our results with those obtained in Refs.~\cite{Hadar:2009ip} and \cite{Hadar:2011vj}, we have displayed the multipolar waveforms and the associated quasinormal ringdowns observed at infinity in the orbital plane of the plunging particle, i.e., for $\theta=\pi/2$. Here, we have only considered the component $h_{+}=h_{+}^{(e)}+h_{+}^{(o)}$ of the emitted gravitational wave (note that  $h_{\times}^{(e)}=h_{\times}^{(o)}=0$) and we have taken for the observation directions in the orbital plane $\varphi=0$, $\varphi=\pi/2$, $\varphi=\pi$ and $\varphi=3\pi/2$, i.e., the angles considered by Hadar and Kol in Fig.~4 of Ref.~\cite{Hadar:2009ip}. We can then realize that the quasinormal ringdowns displayed here do not agree with those of Hadar and Kol. In fact, we can recover their results by plotting the sum $h_{+}^{(e)} - h_{+}^{(o)}$ instead of $h_{+}^{(e)}+h_{+}^{(o)}$ (see Fig.~\ref{HK_hplus_plan_equatorial} and note that the plots in the right panel are in agreement with Fig.~4 of Ref.~\cite{Hadar:2009ip}). Despite a careful study of Ref.~\cite{Hadar:2009ip} and a complete check of our own calculations, we have not been able to identify the cause of this sign difference. Here, it is important to recall that Hadar, Kol, Berti, and Cardoso in Ref.~\cite{Hadar:2011vj} claimed they have confirmed the results of Ref.~\cite{Hadar:2009ip} by comparing them with the Sasaki-Nakamura partial waveforms $X_{\ell m}$. In fact, they have not plotted on a same figure the Sasaki-Nakamura multipolar waveforms and the multipolar quasinormal ringdowns. Their comparison is based on a numerical fitting method which is equivalent to compare the partial modes $|X_{\ell m}|$ with the quasinormal amplitudes $|\psi^{\text{\tiny{QNM}} \, (e)}_{\ell m n}|$ for $\ell + m$ even and $|\psi^{\text{\tiny{QNM}} \, (o)}_{\ell m n}|$ for $\ell + m$ odd. As a consequence, with this method, a wrong sign in the combination of $h_{+}^{(e)}$ and $h_{+}^{(o)}$ cannot be detected.

\section{Conclusion}
\label{Conc}

In this article, we have described the gravitational radiation emitted by a massive ``point particle'' plunging from slightly below the ISCO into a Schwarzschild BH. In order to do this, we have constructed the associated multipolar waveforms and analyzed their late-stage ringdown phase in terms of QNMs. We have noted the excellent agreement between the ``exact'' waveforms we had to carefully regularize and the corresponding quasinormal waveforms which have not required a similar treatment. Our results have been obtained for arbitrary directions of observation and, in particular, outside the orbital plane of the plunging particle. They have permitted us to emphasize more particularly the impact on the distortion of the waveforms of (i) the higher harmonics beyond the dominant $(\ell=2,m=\pm 2)$ modes and (ii) the direction of observation and, as a consequence, the necessity to take them into account in the analysis of the last phase of binary black hole coalescence.

\begin{acknowledgments}

We gratefully acknowledge Thibault Damour for drawing,
some years ago, our attention to the
plunge regime in gravitational wave physics and Shahar Hadar and Barak Kol for useful correspondence. We wish also to thank Yves Decanini and Julien Queva
for various discussions and Romain Franceschini for providing us with powerful computing resources.

\end{acknowledgments}

\appendix*

\section{Regularization of the partial waveform amplitudes (even and odd parity)}
\label{appen}

In this appendix, we shall explain how to regularize the partial amplitudes $\psi_{\omega \ell m}^{(e/o)}$. Indeed, the exact waveforms (\ref{Partial_Response_1}) as integrals over the radial Schwarzschild coordinate are strongly divergent near the ISCO. This is due to the behavior of the sources (\ref{Source_omR_e}) and (\ref{Source_omR_o}) in the limit $r \to 6M$. It should be noted that we have encountered a similar problem in our study of the electromagnetic radiation generated by a charged particle plunging into the Schwarzschild BH \cite{Folacci:2018vtf}. We recall that we have addressed this problem by combining a theoretical and a numerical approach: for even electromagnetic perturbations, we have reduced the degree of divergence of the integrals involved by successive integrations by parts and then we have``numerically regularized'' them by using Levin's algorithm. Here, we can proceed identically and the reader will be assumed to have ``in hand'' a copy of Ref.~\cite{Folacci:2018vtf} where the electromagnetic case is treated in great details in the Appendix. Moreover, we shall use a trick that will allow us to quickly provide the expected results from those obtained in Ref.~\cite{Folacci:2018vtf}.

The trick we use is based on the fact, that the expressions (\ref{Partial_Response_1}), which are constructed from the source terms (\ref{Source_omR_e}) and (\ref{Source_omR_o}), can be written in the form
\begin{equation}\label{Rep_Part_}
\psi_{\omega \ell m}^{(e/o)} (r)=  e^{i \omega r_*} \psi^{(e/o)}_{\ell m}(\omega)
\end{equation}
with
\begin{equation}
\label{Rep_Part_Omega}
\psi^{(e/o)}_{\ell m}(\omega) = \gamma^{(e/o)} \int_{2M}^{6M} dr\, \widetilde{\phi}_{\omega\ell}^{\,\,\mathrm {in} \, {(e/o)}}(r)\widetilde{{\cal A}}^{(e/o)}(r)e^{i \Phi(r)}
\end{equation}
where we have
\begin{equation}
\label{Phase_tot}
\Phi(r) =\omega t_p(r) - m \varphi_p(r)
\end{equation}
and
\begin{subequations}\label{Fact_Gamma}
\begin{eqnarray}
& & \gamma^{(e)} =   \frac{1}{2 i \omega A_\ell^{(-,e)}(\omega)} \frac{8\pi m_0}{\sqrt{2\pi}}  \,\frac{ [Y^{\ell m}(\pi/2,0)]^\ast}{\Lambda+2}, \label{Fact_Gamma_even} \\
& & \gamma^{(o)} =  \frac{1}{2 i \omega A_\ell^{(-,o)}(\omega)} \frac{16\pi m_0}{\sqrt{2\pi}}  \,\frac{ [X_\varphi^{\ell m}(\pi/2,0)]^\ast}{\Lambda(\Lambda+2)}, \label{Fact_Gamma_odd}
\end{eqnarray}
\end{subequations}
and
\begin{equation}
\label{phi_in_tild}
\widetilde{\phi}_{\omega \ell}^{\,\,\mathrm {in} \, {(e/o)}} = \kappa^{(e/o)}(r)\,  \phi_{\omega\ell}^{\mathrm {in} \, {(e/o)}}(r)
\end{equation}
 with
\begin{subequations} \label{Fact_kappa}
\begin{eqnarray}
& & \kappa^{(e)}(r)= - \frac{8}{3\sqrt{2}}\frac{r}{\Lambda r + 6M}, \label{Fact_kappa_even}\\
& & \kappa^{(o)}(r) =\sqrt{3}\left(\frac{2M}{r}\right), \label{Fact_kappa_odd}
\end{eqnarray}
\end{subequations}
as well as
\begin{widetext}
\begin{subequations} \label{Amplitude_tot_evenETodd}
\begin{eqnarray}
& & \widetilde{{\cal A}}^{(e)}(r) =-i\omega \frac{9r(r^2+12 M^2)}{(6M-r)^{3}} +i m \frac{12\sqrt{6}M r}{(6M-r)^{3}}+\frac{18\sqrt{2}M \sqrt{r}}{(6M-r)^{5/2}} + i m\left(\frac{3\sqrt{3}}{2\sqrt{2}}\right)\frac{2M}{r^2} \nonumber \\& & \qquad - \frac{(3\sqrt{2}/8)\sqrt{r}}{(6M-r)^{3/2}}\left[3(\Lambda+2)-\frac{64 M}{\Lambda r+6M}+\frac{72 M^2(\Lambda+2-m^2)}{r^2}+\frac{216 M^3(\Lambda+2-2m^2)}{\Lambda r^3}\right], \label{Amplitude_tot_even} \\
& & \widetilde{{\cal A}}^{(o)}(r) =-i\omega \frac{9r(r^2+12 M^2)}{(6M-r)^{3}} +i m \frac{12\sqrt{6}M r}{(6M-r)^{3}}+\frac{18\sqrt{2}M \sqrt{r}}{(6M-r)^{5/2}} - \frac{4\sqrt{2} \sqrt{r}}{(6M-r)^{3/2}}. \label{Amplitude_tot_odd}
\end{eqnarray}
\end{subequations}
\end{widetext}

We now remark that the amplitudes $\widetilde{{\cal A}}^{(e)}(r)$ and $\widetilde{{\cal A}}^{(o)}(r)$ can be split into a divergent and a regular part in the form
\begin{equation}
\label{Amplitude_tot_bis}
\widetilde{{\cal A}}^{(e/o)}(r)  = \widetilde{{\cal A}}_{\mathrm{div}}(r) + \widetilde{{\cal A}}^{(e/o)}_{\mathrm{reg}}(r)
\end{equation}
and that the divergent part, which is obtained by the Taylor expansion of ${\cal A}^{(e/o)}(r)$ at $r = 6M$, is independent of the parity and given by
 \begin{equation}
\label{Amplitude_div}
\widetilde{{\cal A}}_{\mathrm{div}}(r)  =\frac{c_1}{(6M-r)^3} + \frac{c_2}{(6M-r)^{5/2}} + \frac{c_3}{(6M-r)^2}
\end{equation}
with
\begin{subequations}\label{Coeffs_ci}
\begin{eqnarray}
c_1& =& 18 i\, (2M)^2 \left[\sqrt{6}\, m - 36\, M\omega \right], \\
c_2 &=& 9\sqrt{6}\, (2M)^{3/2},\\
c_3 &=& 6 i\, (2M)  \left[-\sqrt{6}\, m + 90\, M\omega\right].
\end{eqnarray}
\end{subequations}
Here, we fall on the result previously obtained in the context of the regularization of the partial amplitude $\psi_{\omega \ell m}^{(e)}$ describing the electromagnetic radiation generated by a charged particle plunging into the Schwarzschild BH [see Eqs.~(A.6)-(A.8) in Ref.~\cite{Folacci:2018vtf}]. That is a direct consequence of the redefinition (\ref{phi_in_tild})-(\ref{Fact_kappa}) of the functions $\phi_{\omega\ell}^{\mathrm {in} \, {(e/o)}}$. Hence, by noting that the functions $\widetilde{\phi}_{\omega\ell}^{\,\,\mathrm {in} \, {(e/o)}}$  appearing in (\ref{Rep_Part_Omega}) are regular for $r \to 6M$, we can now complete the regularization process by using, {\it mutatis mutandis}, the results obtained in Ref.~\cite{Folacci:2018vtf} and, in particular, Eq.~(A.21) of this article: In order to regularize the partial amplitudes $\psi_{\omega \ell m}^{(e/o)} (r)$ which are given by (\ref{Partial_Response_1}), we therefore replace in Eq.~(\ref{Rep_Part_}) the functions $\psi^{(e/o)}_{\ell m}(\omega)$ by the functions

\begin{widetext}

\begin{eqnarray}\label{REG_RES}
& & \psi^{(e/o) \, \mathrm{reg}}_{\ell m}(\omega)= \gamma^{(e/o)} \int_{2M}^{6M} dr\, {\widetilde{\phi}}_{\omega\ell}^{\,\, \mathrm {in} \, {(e/o)}}(r) \widetilde{{\cal A}}^{(e/o)}_{\mathrm{reg}}(r) e^{i \Phi(r)} \nonumber \\
& & \qquad\qquad\qquad + \frac{3\sqrt{6}}{2}\sqrt{2M}\gamma^{(e/o)}   \left[\int _{2M}^{6M}dr\,\widetilde{\phi}_{\omega \ell}^{\,\, \mathrm {in} \, {(e/o)}}(r) \left(\frac{1}{(6M-r)^{3/2}}+\frac{2id}{(6M-r)}\right) e^{i\Phi(r)}  \right.  \nonumber \\
& & \qquad\qquad\qquad\qquad\qquad\qquad\qquad \left. -2 i\int_{2M}^{6M}dr\, rf(r)\, \frac{\widetilde{\phi}_{\omega \ell}^{\,\, \mathrm {in} \, {(e/o)}}(r) \, \Theta_{\mathrm{reg}}(r)}{(6M-r)^{3/2}} e^{i\Phi(r)}   \right. \nonumber \\
 & & \qquad\qquad\qquad\qquad\qquad\qquad\qquad \left. -2 \int_{2M}^{6M} dr\, rf(r) \frac{\left(\frac{d}{dr}\widetilde{\phi}_{\omega \ell}^{\,\, \mathrm {in} \, {(e/o)}}(r)\right)}{(6M-r)^{3/2}} e^{i\Phi(r)}  \right].
\end{eqnarray}
\end{widetext}
We note that the terms $\widetilde{{\cal A}}^{(e/o)}_{\mathrm{reg}}$ in the r.h.s.~of (\ref{REG_RES}) are obtained from (\ref{Amplitude_tot_evenETodd})-(\ref{Coeffs_ci}). Moreover, the function $\Theta_{\mathrm{reg}}(r)$ is constructed from the phase (\ref{Phase_tot}). This is explained in Appendix of Ref.~\cite{Folacci:2018vtf}. We just recall that
\begin{equation}\label{Theta_reg}
\Theta_{\mathrm{reg}}(r)=\frac{d}{dr}\left[ \Phi(r) - \frac{c}{\sqrt{6M-r}} \right] - \frac{d}{\sqrt{6M-r}}
\end{equation}
where
\begin{subequations}
\begin{equation}
\label{Coeff_c}
c =6\sqrt{2M}\left(m - 6\sqrt{6} \, M\omega \right)
\end{equation}
and
\begin{equation}
\label{Coeff_d}
 d = \frac{m + 12\sqrt{6}  \, M\omega }{2\sqrt{2M}}.
\end{equation}
\end{subequations}

It is finally important to point out that the result (\ref{REG_RES}) has to be ``numerically regularized''. Indeed [see also the discussion in Sec.~(A.2) of Ref.~\cite{Folacci:2018vtf}], the integrands in the r.h.s.~of (\ref{REG_RES}) belong to a particular family of rapidly oscillatory functions whose amplitudes diverge as $1/(6M-r)^{3/2}$ in the limit $r \to 6M$ and whose the phase $\Phi(r)$ behaves as $1/(6M-r)^{1/2}$ in the same limit. As a consequence, it is possible to neutralize the divergences remaining in the amplitudes from the oscillations induced by the phase term. This has been achieved by using Levin's algorithm \cite{Levin1996} which is implemented in {\it Mathematica} \cite{Mathematica}. It is this last step which permits us to obtain, in Sec.~\ref{SecIV}, stable numerical results for the partial amplitudes $\psi_{\omega \ell m}^{(e/o)}(r)$ and $\psi_{\ell m}^{(e/o)}(t,r)$.

\bibliography{Grav_en_S}

\end{document}